\newif\ifpreprint
\preprintfalse 

\ifpreprint
\documentclass[journal=jctcce,manuscript=letter]{achemso}
\else
\documentclass[journal=jctcce,manuscript=letter,layout=twocolumn]{achemso}
\fi

\usepackage[T1]{fontenc} 

\usepackage{amsmath}
\usepackage{newtxtext,newtxmath}
\usepackage{graphicx}
\usepackage{dcolumn}
\usepackage{braket}
\usepackage{multirow}
\usepackage{threeparttable}
\usepackage{xspace}
\usepackage{verbatim}
\usepackage[version=4]{mhchem} 
\usepackage{comment}
\usepackage{color,soul}
\usepackage{multicol}

\usepackage[dvipsnames]{xcolor}
\usepackage{xspace}
\usepackage{ifthen}

\usepackage{qcircuit}

\usepackage{graphicx,longtable,dcolumn,mhchem}
\usepackage{rotating,color}
\usepackage{lscape}
\usepackage{amsmath}
\usepackage{dsfont}
\usepackage{soul}
\usepackage{physics}

\newcommand{\QP}{\textsc{quantum package}\xspace}
\newcolumntype{d}{D{.}{.}{-1}}

\newcommand{\TAVDZ}{t-\textit{aug}-cc-pVDZ}
\newcommand{\TAVTZ}{t-\textit{aug}-cc-pVTZ}
\newcommand{\DAVDZ}{d-\textit{aug}-cc-pVDZ}
\newcommand{\DAVTZ}{d-\textit{aug}-cc-pVTZ}
\newcommand{\DAVQZ}{d-\textit{aug}-cc-pVQZ}

\newcommand{\AVDZ}{\textit{aug}-cc-pVDZ}

\newcommand{\AVTZ}{\textit{aug}-cc-pVTZ}
\newcommand{\AVQZ}{\textit{aug}-cc-pVQZ}

\newcommand{\Td}{\%T_1}

\newcommand{\alltrans}{53} 
\newcommand{\valvaltrans}{9}
\newcommand{\rydvaltrans}{10}
\newcommand{\rydrydtrans}{34}
\newcommand{\allmolecules}{21}
\newcommand{\allstates}{71}
\newcommand{\smallsetmolecules}{5}

\newcommand{\MAEsmallsetCCTHAVDZ}{$0.061$}

\newcommand{\MAEsmallsetCCTHdAVDZ}{$0.008$}

\newcommand{\MAEsmallsetCCTHtAVDZ}{$0.008$}

\newcommand{\MAEsmallsetCCTHAVTZ}{$0.055$}

\newcommand{\MAEsmallsetCCTHAVQZ}{$0.047$}

\newcommand{\MAEallCCTHAVDZ}{$0.020$}
\newcommand{\RMSEallCCTHAVDZ}{$0.047$}
\newcommand{\STEallCCTHAVDZ}{$0.045$}

\newcommand{\MAEallCCTHdAVDZ}{$0.004$}
\newcommand{\RMSEallCCTHdAVDZ}{$0.007$}
\newcommand{\STEallCCTHdAVDZ}{$0.007$}

\newcommand{\MAEallCCTHAVTZ}{$0.017$}
\newcommand{\RMSEallCCTHAVTZ}{$0.042$}
\newcommand{\STEallCCTHAVTZ}{$0.040$}

\newcommand{\MAEallEOMCCSD}{$0.009$}

\newcommand{\MAEallCCSD}{$0.006$}

\newcommand{\MAEallCCTW}{$0.020$}

\newcommand{\MAPEallADCTH}{$0.132$}

\newcommand{\MAEallADCTH}{$0.009$}

\newcommand{\MAEallADCTW}{$0.022$}

\newcommand{\MAEallTDDFTBTHLYP}{$0.036$}

\newcommand{\MAEallTDDFTCAMBTHLYP}{$0.015$}
\newcommand{\RMSEallTDDFTCAMBTHLYP}{$0.037$}
\newcommand{\STEallTDDFTCAMBTHLYP}{$0.036$}

\newcommand{\MAEallTDDFTBHandHLYP}{$0.031$}

\newcommand{\MAEallTDDFTLCBLYPTHTH}{$0.027$}
\newcommand{\RMSEallTDDFTLCBLYPTHTH}{$0.056$}
\newcommand{\STEallTDDFTLCBLYPTHTH}{$0.056$}

\newcommand{\MAPEallTDDFTLCBLYPFOSE}{$0.328$}

\newcommand{\MAEallTDDFTLCBLYPFOSE}{$0.031$}

\newcommand{\MAEallTDABTHLYP}{$0.062$}

\newcommand{\MAEallTDACAMBTHLYP}{$0.044$}

\newcommand{\MAEallTDABHandHLYP}{$0.045$}

\newcommand{\MAEallTDALCBLYPTHTH}{$0.042$}

\newcommand{\MAEallTDALCBLYPFOSE}{$0.044$}

\newcommand{\MAEbigEOMCCSD}{$0.013$}
\newcommand{\RMSEbigEOMCCSD}{$0.018$}
\newcommand{\STEbigEOMCCSD}{$0.013$}

\newcommand{\MAEbigCCTW}{$0.026$}
\newcommand{\RMSEbigCCTW}{$0.036$}
\newcommand{\STEbigCCTW}{$0.031$}

\newcommand{\MAEsmallCCTW}{$0.008$}

\newcommand{\MAEbigADCTW}{$0.029$}
\newcommand{\RMSEbigADCTW}{$0.038$}
\newcommand{\STEbigADCTW}{$0.028$}

\newcommand{\MAEsmallADCTW}{$0.007$}

\newcommand{\RMSEbigTDDFTBTHLYP}{$0.036$}
\newcommand{\STEbigTDDFTBTHLYP}{$0.033$}

\newcommand{\RMSEsmallTDDFTBTHLYP}{$0.048$}
\newcommand{\STEsmallTDDFTBTHLYP}{$0.044$}

\newcommand{\MAEbigTDDFTCAMBTHLYP}{$0.010$}
\newcommand{\RMSEbigTDDFTCAMBTHLYP}{$0.020$}
\newcommand{\STEbigTDDFTCAMBTHLYP}{$0.019$}

\newcommand{\MAEsmallTDDFTCAMBTHLYP}{$0.019$}
\newcommand{\RMSEsmallTDDFTCAMBTHLYP}{$0.053$}
\newcommand{\STEsmallTDDFTCAMBTHLYP}{$0.050$}

\newcommand{\MAEallCCSDEdif}{$0.069$}

\newcommand{\MAEallCCTWEdif}{$0.083$}

\newcommand{\MAEallADCTHEdif}{$0.114$}

\newcommand{\MAEallADCTWEdif}{$0.085$}

\newcommand{\MAEallTDDFTBTHLYPEdif}{$0.351$}

\newcommand{\MAEallTDDFTCAMBTHLYPEdif}{$0.092$}

\newcommand{\MAEallTDDFTBHandHLYPEdif}{$0.082$}

\newcommand{\MAEallTDDFTLCBLYPTHTHEdif}{$0.074$}

\newcommand{\MAEallTDDFTLCBLYPFOSEEdif}{$0.157$}

\newcommand{\MAEallTDABTHLYPEdif}{$0.313$}

\newcommand{\MAEallTDACAMBTHLYPEdif}{$0.080$}

\newcommand{\MAEallTDABHandHLYPEdif}{$0.106$}

\newcommand{\MAEallTDALCBLYPTHTHEdif}{$0.084$}

\newcommand{\MAEallTDALCBLYPFOSEEdif}{$0.243$}

\newcommand{\MAEallADCTWCCTHAVDZ}{$0.037$}

\newcommand{\MAEallADCTWCCTHdAVDZ}{$0.026$}

\newcommand{\MAEallADCTWCCTHAVTZ}{$0.030$}

\newcommand{\MAEallADCTWCCTHdAVTZ}{$0.022$}

\usepackage[colorlinks = true,
            linkcolor = blue,
            urlcolor  = black,
            citecolor = blue,
            anchorcolor = black]{hyperref}

\definecolor{goodorange}{RGB}{225,125,0}
\definecolor{goodgreen}{RGB}{5,130,5}
\definecolor{goodred}{RGB}{220,50,25}
\definecolor{goodblue}{RGB}{30,144,255}

\newcommand{\note}[2]{
\ifthenelse{\equal{#1}{F}}{
\colorbox{goodorange}{\textcolor{white}{\footnotesize \fontfamily{phv}\selectfont #1}}
    \textcolor{goodorange}{{\footnotesize \fontfamily{phv}\selectfont #2}}\xspace
}{}
\ifthenelse{\equal{#1}{R}}{
\colorbox{goodred}{\textcolor{white}{\footnotesize \fontfamily{phv}\selectfont #1}}
    \textcolor{goodred}{{\footnotesize \fontfamily{phv}\selectfont #2}}\xspace
}{}
\ifthenelse{\equal{#1}{N}}{
\colorbox{goodgreen}{\textcolor{white}{\footnotesize \fontfamily{phv}\selectfont #1}}
    \textcolor{goodgreen}{{\footnotesize \fontfamily{phv}\selectfont #2}}\xspace
}{}
\ifthenelse{\equal{#1}{M}}{
\colorbox{goodblue}{\textcolor{white}{\footnotesize \fontfamily{phv}\selectfont #1}}
    \textcolor{goodblue}{{\footnotesize \fontfamily{phv}\selectfont #2}}\xspace
}{}
}

\usepackage{titlesec}

\usepackage[fontsize=11pt]{scrextend}
\titleformat{\section}
{\normalfont\sffamily\bfseries\color{Blue}}
{\thesection.}{0.25em}{\uppercase}

\titleformat{\subsection}
{\normalfont\sffamily\bfseries}
{\thesubsection}{0.25em}{}

\titleformat{\subsubsection}
{\normalfont\sffamily}
{\thesubsubsection}{0.25em}{}

\titleformat{\suppinfo}
{\normalfont\sffamily\bfseries}
{\thesubsection}{0.25em}{}

\titlespacing*{\section}{0pt}{0.5\baselineskip}{0.01\baselineskip}
\titlespacing*{\subsection}{0pt}{0.125\baselineskip}{0.01\baselineskip}
\titlespacing*{\subsubsection}{0pt}{0.125\baselineskip}{0.01\baselineskip}

\author{Jakub \v{S}ir\r{u}\v{c}ek}
	\affiliation[CEISAM, Nantes]{Nantes Universit\'e, CNRS, CEISAM UMR 6230, F-44000 Nantes, France}
	\alsoaffiliation[ISCR, Rennes]{Univ Rennes, CNRS, ISCR UMR 6226, F-35042 Rennes, France}
\author{Boris Le Guennic}
	\affiliation[ISCR, Rennes]{Univ Rennes, CNRS, ISCR UMR 6226, F-35042 Rennes, France}
        \email{Boris.Leguennic@univ-rennes.fr}
\author{Yann Damour}
              \affiliation[Laboratoire de Chimie et Physique Quantiques, Toulouse]{Universit\'e de Toulouse, CNRS, UPS, Laboratoire de Chimie et Physique Quantiques UMR 5626, F-31062 Toulouse, France}
\author{Pierre-Fran\c cois Loos}
        \email{loos@irsamc.ups-tlse.fr}
        \affiliation[Laboratoire de Chimie et Physique Quantiques, Toulouse]{Universit\'e de Toulouse, CNRS, UPS, Laboratoire de Chimie et Physique Quantiques UMR 5626, F-31062 Toulouse, France}
\author{Denis Jacquemin}
        \email{Denis.Jacquemin@univ-nantes.fr }
	\affiliation[CEISAM, Nantes]{Nantes Universit\'e, CNRS, CEISAM UMR 6230, F-44000 Nantes, France}
        \alsoaffiliation[Institute Universitaire de France (IUF)]{Institut Universitaire de France (IUF), F-75005 Paris, France}

\let\oldmaketitle\maketitle
\let\maketitle\relax
	\title{Excited State Absorption: Reference Oscillator Strengths, Wavefunction and TD-DFT Benchmarks}
\date{\today}

\begin{tocentry}
    \centering
     \includegraphics[width=3.5cm]{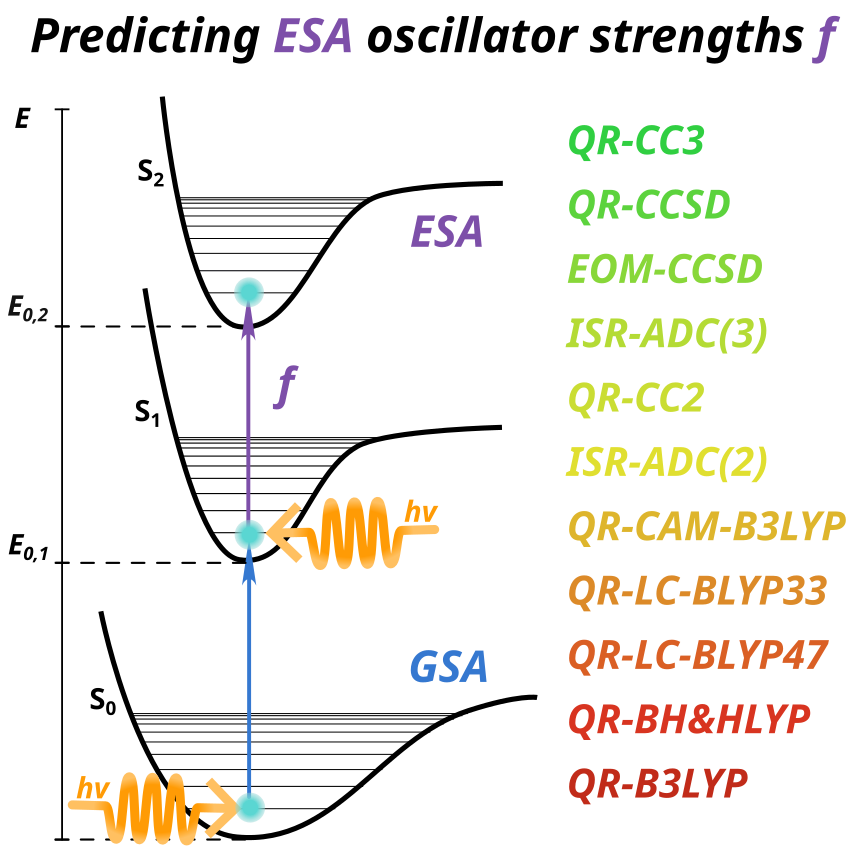} 
\end{tocentry}

\begin{document}

\ifpreprint
\else
\twocolumn[
\begin{@twocolumnfalse}
\fi
\oldmaketitle
\begin{abstract}
Excited-state absorption (ESA) corresponds to the transition between two electronic excited states and is a fundamental process for probing and understanding light-matter interactions. Accurate modeling of ESA is indeed often required to interpret time-resolved experiments. In this contribution, we present a dataset of {\alltrans} ESA oscillator strengths in three different gauges and the associated vertical transition energies between {\allstates} excited states of {\allmolecules} small- and medium-sized molecules from the QUEST database. The reference values were obtained within the quadratic-response (QR) CC3 formalism using eight different Dunning basis sets. We found that the {\DAVTZ} basis set is always adequate while its more compact double-$\zeta$ counterpart, {\DAVDZ}, performs well in most applications. These QR-CC3 data allow us to assess the performance of QR-TDDFT, with and without applying the Tamm-Dancoff approximation, using a panel of global and range-separated hybrids (B3LYP, BH{\&}HLYP, CAM-B3LYP, LC-BLYP33, and LC-BLYP47), as well as several lower-order wavefunction methods, i.e., QR-CCSD, QR-CC2, EOM-CCSD, ISR-ADC(2), and ISR-ADC(3). We show that QR-TDDFT delivers acceptable errors for ESA oscillator strengths, with CAM-B3LYP showing particular promise, especially for the largest molecules of our set. We also find that ISR-ADC(3) exhibits excellent performance.
\end{abstract}

\ifpreprint
\else
\end{@twocolumnfalse}
]
\fi

\ifpreprint
\else
\small
\fi

\noindent

\section{Introduction}

During a typical ground-state absorption (GSA), the electronic cloud is excited from its ground state to an excited state (ES) by absorbing a photon. 
Excited-state absorption (ESA) is a similar process wherein the electronic transition occurs between two ESs. In transient spectroscopies, ESA stands as a crucial photophysical phenomenon for probing and understanding light-matter interactions. \cite{jailaubekov_hot_2013}

ESA is key to a wide range of technological applications from solar cells, \cite{gunes_conjugated_2007} lasers, \cite{samuel_organic_2007} and light-emitting diodes, \cite{tang_organic_1987} to chemical sensors, \cite{thomas_chemical_2007} optical amplifiers, \cite{amarasinghe_high-gain_2009} and optical power-limiting devices. \cite{babeela_excited_2019, bellier_excited_2012, han_synthesis_2022} In these applications, transitions between ESs play a pivotal role in device operation and are parts of the many complex ESs processes occurring in, e.g., exciton-exciton annihilation, \cite{shaw_exciton_2008} exciton-polaron quenching, \cite{yamamoto_amplified_2004} electric-field-induced ionization of excitons, \cite{gartner_numerical_2007} reabsorption of emitted light by polarons, \cite{brown_optical_1993} and two-photon absorption. \cite{bellier_excited_2012,han_synthesis_2022} 

Illustratively, combining ESA with two-photon absorption in optical power-limiting devices allows for maintaining transparency at low light intensities while achieving increased absorption at higher intensities, paving the way to eye-protecting devices. This approach has been demonstrated in several molecular architectures such as organic and organometallic cyanines \cite{bellier_excited_2012} and twistacenes. \cite{han_synthesis_2022} 

Accurate reference ESA data are thus increasingly sought after for their relevance in designing improved molecules and materials. Moreover, in many experiments, ESA cannot be easily separated from other photoinduced signatures contributing to the spectral properties in the same energy range, and thus accurate calculations of ESA are helpful to distinguish ESA from other phenomena. \cite{tautz_structural_2012, brown_optical_1993, tautz_charge_2013}

When computing ESA properties, two aspects ought to be considered: the energy difference, often expressed as the difference of vertical excitation energies (${\Delta}E_{m,n}$) between the $m$th and $n$th ESs, and the transition probability, which can be expressed in terms of transition dipole moments ($\mu$) or oscillator strengths ($f$), and which are typically more demanding to determine.

To obtain $f$ values for ESA, or GSA, one has to select an appropriate electronic structure method. Among the most widely used ones for ESA are: various flavors of coupled-cluster (CC) theory, such as CC2, \cite{christiansen_second-order_1995,cronstrand_ab_2001, cronstrand_theoretical_2000} and CCSD; \cite{koch_coupled_1990,purvis_full_1982,cronstrand_ab_2001,cronstrand_theoretical_2000, fedotov_excited-state_2021, fedotov_excited_2022} multireference (MR) approaches, most notably CASPT2 (second-order complete-active-space perturbation theory); \cite{cronstrand_theoretical_2000,roldao_accurate_2022,foggi_s_2003} and, of course, time-dependent density-functional theory (TD-DFT). \cite{roldao_accurate_2022,parker_quadratic_2018, fedotov_excited-state_2021, fischer_excited_2015, fedotov_excited_2022, bowman_excited-state_2017, ling_excited-state_2013, zhu_linear_2021, roseli_origin_2017, sheng_excited-state_2020, de_wergifosse_nonlinear-response_2019} The latter is computationally cheap and remains the most widely employed in practice. However, TD-DFT results significantly depend on the selected exchange-correlation functional (XCF), often leading to unanswered questions regarding the accuracy that can be expected from TD-DFT.   

CC3 \cite{cronstrand_theoretical_2000, fedotov_excited_2022, cronstrand_ab_2001, fedotov_excited-state_2021} is an iterative CC formalism that includes single, double, and triple excitations. \cite{koch_cc3_1997} It has been shown to deliver highly accurate ES estimates with, in particular, ${\Delta}E_{0,n}$ ($m=0$ being the GS) and GSA oscillator strengths that can serve as solid references for benchmark studies. \cite{sarkar_benchmarking_2021, loos_mountaineering_2018, schreiber_benchmarks_2008} Hence, we rely on CC3 to establish our reference ESA values in the present work.

More globally, the CC approach is, in general, more user-friendly than MR methods as it requires virtually no knowledge about the system and its corresponding ESs. It is thus a preferable choice when dealing with ESs possessing a dominant single-reference (SR) character in a low-coupling region. \cite{cronstrand_theoretical_2000} Additionally, the accuracy of CC-based $f$ values can be probed by gauge invariance. \cite{pawlowski_gauge_2004, pedersen_gauge_1998} Indeed, the oscillator strengths can be computed in the length, velocity, and mixed gauges, as follows:
\begin{subequations}
\begin{align}
    f_{m,n}^{\mathrm{l}} & = \frac{2\Delta{E}_{m,n}}{3}\mel{m}{\textbf{r}}{n}\mel{n}{\textbf{r}}{m} 
    \label{eq:length}
    \\
    f_{m,n}^{\mathrm{v}} & = \frac{2}{3\Delta{E}_{m,n}}\mel{m}{\textbf{p}}{n}\mel{n}{\textbf{p}}{m} 
    \label{eq:velocity}
    \\
    f_{m,n}^{\mathrm{m}} & = -\frac{2i}{3}\mel{m}{\textbf{r}}{n}\mel{n}{\textbf{p}}{m} 
    \label{eq:mixed}
\end{align}
\end{subequations}
where $f$ with superscripts l, v, and m denote the oscillator strengths in length, velocity, and mixed gauge, respectively, of the transition between the $m$th and $n$th ESs, \textbf{r} is the length operator and \textbf{p} the momentum operator. As a reminder, we denote the GSA vertical excitation energy as ${\Delta}E_{0,n}$ and the corresponding ESA energy as ${\Delta}E_{m,n}$. For the exact wavefunction, that is, the full configuration interaction (FCI) wavefunction computed in a complete basis set, these three equations deliver the same results, illustrating the expected gauge invariance for the exact wavefunction. \cite{pedersen_coupled_1997} Hence, when one considers an approximate level of theory, inspecting the difference between the three gauges is a crude metric of the accuracy of the calculated $f$ values.
Indeed, it was shown by Pawlowski \textit{et al.}, \cite{pawlowski_gauge_2004} using the CCS, CC2, CCSD, and CC3 series, and later by some of us, \cite{sarkar_benchmarking_2021} that the difference between GSA oscillator strengths obtained in different gauges becomes markedly smaller when the maximum excitation degree of the cluster operator increases.
Of course, when the desired ESs exhibit MR characters, SR approaches become less efficient and one has to rely on MR schemes such as the popular CASPT2 formalism. \cite{roldao_accurate_2022} MR methods can, in principle, accurately describe all ESs but can be challenging to use. \cite{roldao_accurate_2022, cronstrand_theoretical_2000, malmqvist_casscf_1989}

At this stage, we recall that there are two main CC approaches for computing ES properties: the equation-of-motion (EOM)\cite{stanton_equation_1993} and response function formalisms, the latter being also known as response theory (RT).\cite{koch_coupled_1990, pedersen_coupled_1997} In the RT approach, time-independent expectation values of molecular properties are expanded in orders of a frequency-dependent perturbation leading to results equivalent to those reached with numerical derivatives. In the EOM formalism, molecular properties are computed directly from the expectation value of the corresponding operator for the physical observable. It should be noted that both approaches yield the same vertical excitation energies but differ in transition dipoles. Nevertheless, at high levels of theory, such as CC3, the computationally less expensive EOM gives GSA $f$ values very similar to the ones obtained with RT. \cite{paul_new_2021, bartlett_coupled-cluster_2012, koch_calculation_1994, caricato_difference_2009, sarkar_benchmarking_2021}

Apart from RT and EOM, we wish to mention the intermediate state representation (ISR) which has been extensively used within the ADC (algebraic diagrammatic construction) family of methods, \textit{e.g.} ADC(2) and ADC(3). \cite{schirmer_beyond_1982, dreuw_algebraic_2015} 
In this work, we primarily focus on the RT approach, which has been widely used for GSA and, in several cases, for ESA calculations as well, and is readily applicable within TD-DFT. \cite{roldao_accurate_2022, parker_quadratic_2018, ling_excited-state_2013, zhu_linear_2021, cronstrand_theoretical_2000} 

In the RT framework, there are, in principle, two options for obtaining the ESA $f$ values: from the single residues of the linear response (LR) function of one of the two ESs involved in the ESA or from the double residues of the quadratic response (QR) function of the ground state. \cite{cronstrand_theoretical_2000, han_synthesis_2022} The latter formalism allows not only calculating the ESA oscillator strengths but also two-photon absorption properties which are often sought after in combination with ESA, \cite{bellier_excited_2012, han_synthesis_2022} as discussed above. 

To the best of our knowledge, the first seminal study testing different RT approaches on ESA calculations was performed by Norman \textit{et al.}~in 1996. \cite{norman_excited_1997} Both SR and MR models were tested with cubic and linear RTs on the ESs of benzene and naphthalene. With cubic response theory, one can obtain ES polarizabilities in a similar fashion as computing $f$ values with QR. For the ES polarizabilities, the authors found a good performance of the cubic response theory, which in some cases outperformed the LR approach.  

Subsequent studies by Cronstrand \textit{et al.}~in 2000\cite{cronstrand_theoretical_2000} and 2001\cite{cronstrand_ab_2001} further explored ESA with different methods. In the former work, the authors compared the QR and LR with a hierarchy of CC methods [CCS, CC2, CCSD, CCSDR(3), and CC3] on small molecules, confirming the superiority of the QR approach. In the second study, the QR approach was tested with the same hierarchy of CC methods on butadiene and a set of polyenes. Both works showed a very good agreement of CC3 $f$ values and ${\Delta}E_{0,n}$ with experimental data when a triple-$\zeta$ basis set is employed. Moreover, a systematic convergence of $\mu$ in the CC series was reported.

We underline that in the case of CC3 and CCSDR(3), both studies computed the $f$ values using CCSD $\mu$ values and the corresponding CC3 or CCSDR(3) energies. This approach was necessary because, on the one hand, CCSDR(3) does not allow for calculating $\mu$ values within the RT formalism, owing to its non-iterative, perturbative treatment of triple excitations, \cite{christiansen_perturbative_1996} while, on the other hand, CC3 theoretically permits such calculations, albeit no implementation was available at the time.   

Of course, most wavefunction approaches are computationally expensive, which poses a significant limitation, especially in ESA applications where the molecules of interest are typically organic or organometallic dyes, sometimes with an oligo or polymeric character. \cite{gunes_conjugated_2007, samuel_organic_2007, tang_organic_1987, amarasinghe_high-gain_2009, thomas_chemical_2007, babeela_excited_2019, bellier_excited_2012} For these reasons, QR-TDDFT is a cheap and useful alternative to wavefunction approaches.

The first work, to the best of our knowledge, utilizing QR-TDDFT for ESA calculations, was published by Ling \textit{et al.}~in 2013 \cite{ling_excited-state_2013}. In this study, QR-TDDFT was applied to model near-infrared (NIR) spectra of oligofluorenes, molecules that are considerably larger than those typically accessible by wavefunction-based methods. The authors demonstrated a good qualitative agreement with the measured ESA spectra corresponding to transitions from the $S_1$ state. Subsequently, the performance of QR-TDDFT was compared with its linear counterpart using the same set of oligofluorenes. \cite{zhu_linear_2021} While the computationally cheaper LR model \cite{sheng_excited-state_2020, fischer_excited_2015, fischer_excited-state_2016} provided energies similar to those obtained with QR in the NIR region, significant variations were observed for the $f$ values. Both studies were limited to a dataset of 7 oligofluorenes and mainly focused on transitions from the $S_1$ state. A wider set of fluorene homo- and co-polymers was investigated by Denis \textit{et al.}~in 2016 \cite{denis_self-trapping_2016} within the Tamm-Dancoff approximation (TDA) of QR-TDDFT. Their findings also showed good qualitative agreement with the experimental data in the NIR region. We underline that these studies revealed minimal sensitivity of the ESA spectra to geometry relaxation from the Franck-Condon geometry to the $S_1$ minima. Following these initial investigations, the amount of ESA QR-TDDFT calculations has grown over the last few years. \cite{parker_quadratic_2018, fedotov_excited-state_2021, fedotov_excited_2022, bowman_excited-state_2017, de_wergifosse_nonlinear-response_2019} While consistent qualitative agreement has been achieved for energetically low-lying ESA transitions, \cite{han_evaluation_2020, roseli_origin_2017} challenges persist for higher-lying transitions. \cite{vishal_first-principles_2020}. One of the significant obstacles for capturing the higher-lying ESA transitions is state characterization, as pinpointed by Roldao \textit{et al.} \cite{roldao_accurate_2022} 

Despite the availability of benchmarks and datasets for GSA, such as the QUEST, \cite{QUESTDB_span_2021} VERDE, \cite{abreha_virtual_2019} and QM-symex \cite{ling_excited-state_2013} databases, reference ESA data remain scarce. A comprehensive benchmark covering a wide range of $f$ values, different methods, XCFs (for TD-DFT), and basis sets for ESA is lacking. Here, we present such a dataset comprising $\alltrans$ ES transitions in $\allmolecules$ molecules containing from 1 to 6 (non-hydrogen) atoms. The selected transitions occur between energetically low-lying Rydberg and valence ESs with a predominant single excitation character.

Below, we first investigate the convergence of $f$ values at the QR-CC3 level to establish computationally attainable yet trustworthy reference values. For this purpose, we compute the values of the oscillator strengths in all three gauges (length, velocity, and mixed) with eight different correlation-consistent basis sets varying in the level of augmentation with diffuse functions (single to triple) and $\zeta$-multiplicity (double to quadruple) on a set of small molecules. To further confirm our choice of {a} reference level of theory, we conduct extrapolated FCI calculations. Next, we conduct the remaining part of the QR-CC calculations in four basis sets, which we also employ to assess the ISR implementation of the second-order ADC [ISR-ADC(2)], \cite{dreuw_algebraic_2015} which was shown to yield results of accuracy similar to CC2 for GSA. \cite{sarkar_benchmarking_2021,loos_mountaineering_2018} For the rest of our study, we utilize the selected reference {\DAVTZ} basis set only and evaluate various wavefunction approaches [ISR-ADC(3) and EOM-CCSD] as well as QR-DTDFT, with and without TDA. We choose two global hybrids, B3LYP, \cite{becke_density-functional_1993, lee_development_1988, stephens_ab_1994, vosko_accurate_1980} BH\&HLYP, \cite{becke:jcp:1993} and three range-separated hybrids, namely, CAM-B3LYP, \cite{yanai_new_2004} and two different versions of LC-BLYP. \cite{iikura_long-range_2001} The two versions vary in the range separation parameter $\omega$ and we refer to them as LC-BLYP33 ($\omega = 0.33$ $\mathrm{bohr}^{-1}$) and LC-BLYP47 ($\omega = 0.47$ ${\mathrm{bohr}^{-1}}$) 

We anticipate that the comprehensive dataset generated in this study will be helpful in the field of ESA calculations and will hopefully contribute to the design and development of novel functional materials. 

  \begin{figure*}[htp]
   \centering
	    \includegraphics[width=.8\linewidth]{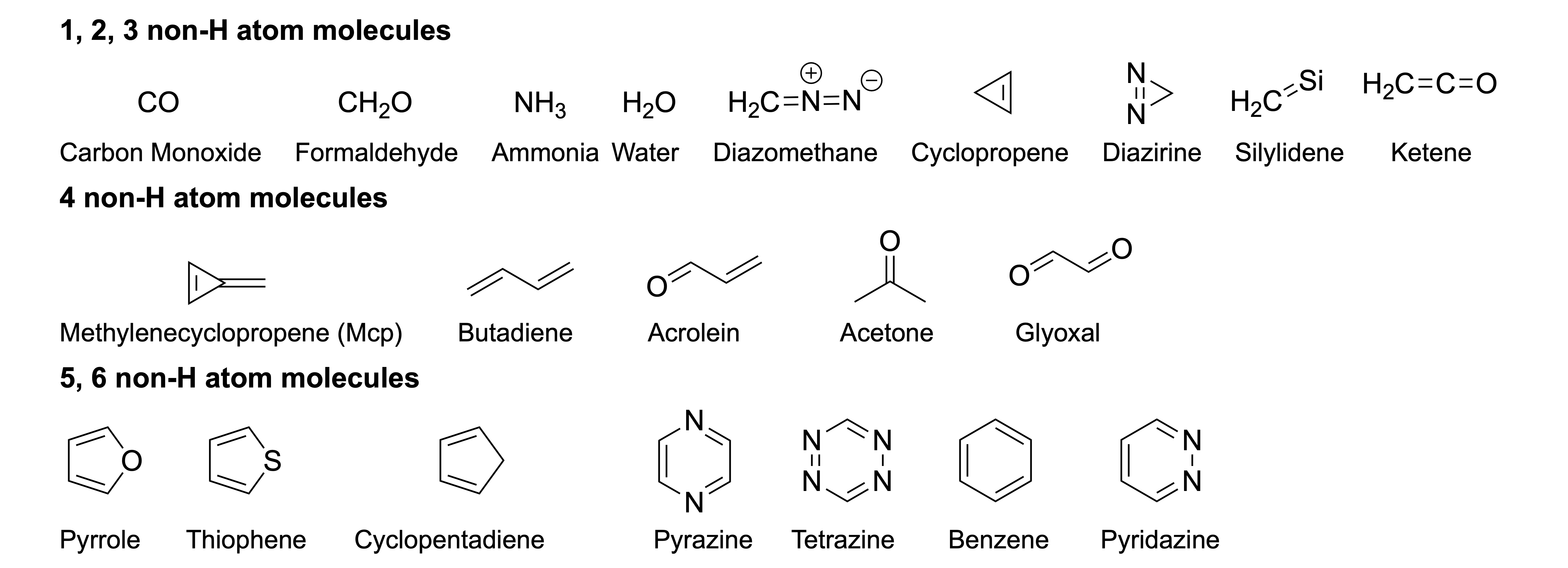}
   \caption{Schematic representation of the molecules considered in this study.}
   
	   \label{fig:Figure-1}
    \end{figure*}

\section{Computational details}
\label{sec:comp_det}

The selected set of {\allmolecules} molecules is shown in Fig.~\ref{fig:Figure-1}. All calculations were performed using the CC3 ground-state geometries of the QUEST database, \cite{QUESTDB_span_2021} which are provided in the supporting information (SI). These computations employ the frozen-core (FC) approximation, except for the TD-DFT calculations.
Unless otherwise noted, the default convergence thresholds and algorithms of the selected codes were used.

All QR-CC calculations (CC2, CCSD, and CC3) of $f$ values in all three gauges were done using the double residues of the quadratic response function within the approach implemented in Dalton 2020. \cite{Dalton, daltonpaper, pecul_high-order_2006, hattig_frequency-dependent_1997} All QR-CC $f$ values were taken directly from the Dalton output except for the mixed gauge. While computing $f$ in the mixed gauge, Dalton often prints the correct absolute value of $f$ but with a negative sign. We suspect that this is a misprint. When the value of $f$ is recomputed from the transition strengths, printed in the Dalton output, as expressed in Eq.~\eqref{eq:mixed}, it indeed yields the same absolute value with the correct (positive) sign. Therefore, in such cases, we only report the absolute $f$ values.

QR-CC calculations were first conducted on a small set of transitions in compact molecules: ammonia, carbon monoxide, diazirine, silylidene, and water with a total of eight Dunning augmented correlation consistent basis sets, as implemented in Dalton: {\AVDZ} (AVDZ), {\DAVDZ} (dAVDZ), {\TAVDZ} (tAVDZ), {\AVTZ} (AVTZ), {\DAVTZ} (dAVTZ), {\TAVTZ} (tAVTZ), {\AVQZ} (AVQZ), {\DAVQZ} (dAVQZ). The dAVTZ basis set was deemed sufficiently complete for our purposes with an acceptable computational cost and was thus employed to establish our reference values (\textit{vide infra}). After this preliminary examination, the rest of the QR-CC and ADC(2) calculations were performed with the AVDZ, dAVDZ, AVTZ, and dAVTZ basis sets only. 

All ESA ADC(2) and ADC(3) values were obtained using ISR as implemented in Q-Chem 6.0. {\cite{schirmer_intermediate_2004, Qchem6}} The SCF convergence threshold was set to at least $10^{-8}$ a.u. When the requested ESs did not converge, the default number of iterations in the Davidson procedure was increased. The same four basis sets as above were used for ADC(2), each combined with the appropriate auxiliary basis, as defined in Q-Chem. On top of that, a smaller secondary basis set (cc-pVTZ) was used for the initial guess at the beginning of all AVTZ and dAVTZ calculations. In the case of ADC(3), we used only the dAVTZ basis set. In a few cases, when the target molecule possesses high symmetry, the program could not determine the irreducible representations of the orbitals. In such cases, the calculations were performed without the secondary initial guess basis set. Only length gauge $f$ values were obtained with ADC(2). 

All EOM-CCSD calculations were done using Q-Chem 6.0. {\cite{stanton_equation_1993}} We ensured the correct state assignment by comparison with QR-CCSD vertical excitation energies ${\Delta}E_{0,n}$, maintaining an acceptable numerical error of 0.01 eV. The calculated $f$ values were verified to correspond to transitions from the lower-energy ES to the higher-energy one. The dAVTZ basis set was considered. Similar to the ADC(2) and ADC(3) calculations, the appropriate auxiliary basis set was used alongside a smaller secondary basis set. However, the latter was occasionally omitted when molecular symmetry constraints prevented its application. Only length-gauge $f$ values were obtained with EOM-CCSD. 

We tested five different XCFs: two global hybrids, namely B3LYP{\cite{becke_density-functional_1993, lee_development_1988, stephens_ab_1994, vosko_accurate_1980}} (20{\%} of exact exchange), and BH\&HLYP{\cite{becke:jcp:1993}} (50{\%} of exact exchange), and three range-separated hybrids, namely CAM-B3LYP{\cite{yanai_new_2004}}, with exact exchange ranging from 19{\%} to 65{\%} and a range separation parameter $\omega$ of 0.33 bohr$^{-1}$, and two different versions of LC-BLYP{\cite{iikura_long-range_2001}} for which exact exchange ranges from 0{\%} to 100{\%}. The two versions vary in the range separation parameter $\omega$ and we refer to them as LC-BLYP33 ($\omega = 0.33$ bohr$^{-1}$) and LC-BLYP47 ($\omega = 0.47$ bohr$^{-1}$).
All ESA TD-DFT calculations, with and without TDA, of ${\Delta}E_{m,n}$ and $f$, in length gauge, were done using the double residues of the quadratic response function as implemented in Dalton 2020. {\cite{jorgensen_linear_1988, vahtras_multiconfigurational_1992, agren_direct_1993}} Only the dAVTZ basis set was used as implemented in Dalton. A SCF convergence threshold of at least $10^{-8}$ a.u.~was used. The transition dipoles in the length and velocity gauges were obtained as the norm of the corresponding $x$, $y$, and $z$ transition dipole moments printed in the Dalton output. 
The final $f$ values were then computed according to Eqs.~\eqref{eq:length} and \eqref{eq:velocity}.

Extrapolated FCI values were derived from CIPSI calculations performed using the \QP software. \cite{Garniron_2019}
For each transition, the two states of interest were identified based on EOM-CCSD results and extracted from the CIS wave function. 
CIPSI calculations were carried out for each pair of states using the state-following procedure employed in Ref.~\citenum{Damour_2024}. 
The extrapolated FCI values and associated uncertainties were then obtained through a 4-point weighted quadratic fit, following the procedure explained in Ref.~\citenum{Damour_2024}.
Since two states were involved, the extrapolations rely on the average of their respective second-order energies, as detailed in Ref.~\citenum{Damour_2023}.

\section{Results and discussion}

To analyze our data, we employ six statistical indicators: the mean signed error (MSE), the mean absolute error (MAE), the root mean square error (RMSE), the standard deviation of errors (SDE), the mean signed percentage error (MSPE), and the mean absolute percentage error (MAPE), defined as follows:
\begin{subequations}
\begin{align}
    \mathrm{MSE} & = \frac{1}{N}\sum_{i=1}^{N}{(y_{i}^{\mathrm{cur.}} - y_{i}^{\mathrm{ref.}})} 
    \label{eq:MSE}
    \\
    \mathrm{MAE} & = \frac{1}{N}\sum_{i=1}^{N}{|y_{i}^{\mathrm{cur.}} - y_{i}^{\mathrm{ref.}}|}  \label{eq:MAE}
    \\
    \mathrm{RMSE} & = \sqrt{\frac{1}{N}\sum_{i=1}^{N}{(y_{i}^{\mathrm{cur.}} - y_{i}^{\mathrm{ref.}})}^2} \label{eq:RMSE}
    \\
    \mathrm{SDE} & = \sqrt{\frac{1}{N}\sum_{i=1}^{N}{(y_{i}^{\mathrm{cur.}} - {\mathrm{MSE}})}^2} \label{eq:STE}
    \\
    \mathrm{MSPE} & = \frac{1}{N}\sum_{i=1}^{N} 
    {\frac{(y_{i}^{\mathrm{cur.}} - y_{i}^{\mathrm{ref.}})}{y_{i}^{\mathrm{ref.}}}} \label{eq:MSPE}
    \\
    \mathrm{MAPE} & = \frac{1}{N}\sum_{i=1}^{N} 
    {\frac{|y_{i}^{\mathrm{cur.}} - y_{i}^{\mathrm{ref.}}|}{y_{i}^{\mathrm{ref.}}}}
    \label{eq:MAPE}
\end{align}
\end{subequations}
where $y_{i}^{\mathrm{cur.}}$ and $y_{i}^{\mathrm{ref.}}$ are the current and reference values of the $i$th transition/state, respectively, and {$N$} is the total number of transitions/states.

  \begin{figure*}[btp]
   \centering
	    \includegraphics[width=.7\linewidth]{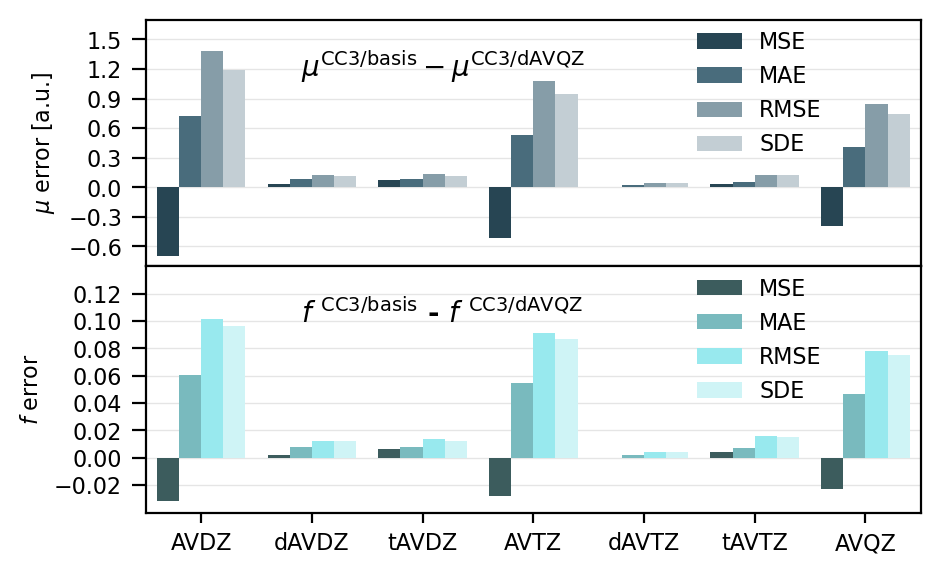}
   \caption{MSE, MAE, RMSE, and SDE of $\mu$ (top) and $f$ in length gauge (bottom), computed with seven different basis sets (AVDZ, dAVDZ, tAVDZ, AVTZ, dAVTZ, tAVTZ, and AVQZ) sets and QR-CC3. QR-CC3/dAVQZ is used as a reference. }
	   \label{fig:Figure-2}
    \end{figure*}
    
The majority of transitions considered here are of predominantly single excitation character ($\Td > 85\%$), based on our reference CC3/dAVTZ calculations. The few states with $\Td < 85\%$ are listed in Table \ref{tab:Table-1}. While CC3 is likely less accurate for these states, $\Td$ remains rather high in all cases. All states are fully characterized in the SI.

\begin{table}
\small
   \begin{tabular}{ l @{\hspace{10pt}} l @{\hspace{10pt}} c @{\hspace{10pt}} c @{\hspace{8pt}} } 
   \hline
Molecule & State & ${\Delta}E_{0,n}$ & $\Td$ \\ 
\hline
      Acrolein    & 2$\mathrm{A}''$ (V, {n-$\pi^*$})    & 6.743    & 79.5  \\
Cyclopentadiene   & 2$\mathrm{A}_1$ (V, {$\pi$-$\pi^*$})    & 6.567    & 78.3  \\
        Glyoxal   & 2$\mathrm{B}_{\mathrm{g}}$ (V, {n-$\pi^*$})    & 6.566    & 84.0  \\
      Tetrazine   & 1$\mathrm{B}_{\mathrm{1g}}$ (V, {n-$\pi^*$})    & 4.910    & 83.1  \\
                  & 1$\mathrm{B}_{\mathrm{2g}}$ (V, {n-$\pi^*$})    & 5.463    & 81.8  \\
        \hline
\end{tabular}
    \caption{States included in this study exhibiting relatively low single excitation character ($\Td$ < 85\%). Both energy (in eV) and $\Td$ were obtained from our reference CC3/dAVTZ calculations. All states have a valence (V) character.} 
    \label{tab:Table-1}
\end{table}                 

A challenge encountered when computing reference values is state characterization as, in some cases, the orbital composition of a given state may change significantly with the basis set extension. To assign unambiguously the computed ES, we initially calculated CC3/AVDZ values and compared them with the QUEST database, followed by CC3/dAVTZ values, which were compared to the AVDZ results. In cases where assignment was not possible based solely on GSA $f$ and vertical excitation energy (${\Delta}E_{0,n}$), it was necessary to analyze the orbital composition. Occasionally, some states became too mixed when going from AVDZ to dAVTZ, or \textit{vice versa}. This issue was particularly common for energetically higher-lying states. For example, the $1{\Pi}_{u}$ state of \ce{N2} and the $3\mathrm{A}_1$ state of formaldehyde exhibit significant state mixing upon removal/addition of diffuse functions. Transitions involving such mixed states were excluded from our set, as their character was too ambiguous for a reliable characterization.

Some molecules have a non-Abelian point group symmetry. The majority of quantum chemistry program packages, including the ones we employed, cannot perform calculations in a non-Abelian point group. In these cases, the program descends to a lower-symmetry point group, that is, the highest-symmetry Abelian point group. This can lead to arbitrary degeneracies when an irreducible representation of the original non-Abelian point group corresponds to two possible irreducible representations in the new lower-symmetry Abelian point group. The two degenerate states have the same $f$, $\mu$, and ${\Delta}E_{\mathrm{0,n}}$. To obtain the $f$ ($\mu$) values of the original state, one has to sum up the corresponding degenerate values of $f$ ($\mu$), i.e., multiply them by two. In the case of ESA, if both states involved in the transition are degenerate, one has to multiply the corresponding value of $f$ ($\mu$) by four. In this work, we publish $f$ ($\mu$) of only one of the degenerate states/transitions, and we highlight the degenerate states with an asterisk (*).

\subsection{Reference values}

To select an appropriate reference basis set, we study a small set of transitions provided in Table \ref{tab:Table-2} using the very large {\DAVQZ} basis. All states taking part in these transitions have a predominantly single excitation character ($\Td > 85\%$). The MSE, MAE, RMSE, and SDE of $\mu$ and $f$ computed with different basis sets at the QR-CC3 level are given in Fig.~{\ref{fig:Figure-2}}. The data clearly underscore the importance of double augmentation on both $f$ and $\mu$ values, as well as the very similar error patterns for these two properties. 

\begin{table}
\small
   \begin{tabular}{ l @{\hspace{15pt}} l @{\hspace{10pt}} c @{\hspace{8pt}} c @{\hspace{8pt}} c }
        \hline
        Molecule & Initial state & Final state  & ${\Delta}E_{m, n}$ &   $f$  \\
        \hline
Ammonia & 1$\mathrm{A}_2$ (R)  & 1E (R)* & 1.547 & 0.319   \\
 & 1$\mathrm{A}_2$ (R)  & 2$\mathrm{A}_1$ (R)  &  1.995 & 0.294 \\
 & 1E (R)* & 2$\mathrm{A}_2$ (R)  &  1.006 & 0.134 \\
 & 2$\mathrm{A}_1$ (R)  & 2$\mathrm{A}_2$ (R)  &  0.558 & 0.264 \\
CO & 2$\mathrm{\Sigma}^{+}$ (R)  & 3$\mathrm{\Sigma}^{+}$ (R)  & 0.609 & 0.146   \\
 & 2$\mathrm{\Sigma}^{+}$ (R)  & 2$\mathrm{\Pi}$ (R)* &  0.737 & 0.255 \\
 & 2$\mathrm{\Pi}$ (R)*  & 1$\mathrm{\Pi}$ (V)* & 2.986 & 0.010   \\
Diazirine & 1$\mathrm{B}_1$ (V)   & 2$\mathrm{A}_1$ (V) & 3.909 & 0.014   \\
 & 1$\mathrm{B}_2$ (R)   & 2$\mathrm{A}_1$ (V)  & 0.537 & 0.080   \\
Silylidene & 1$\mathrm{A}_2$ (V)   & 1$\mathrm{B}_2$ (V) & 1.626 & 0.003   \\
Water & 1$\mathrm{B}_1$ (R)  & 1$\mathrm{A}_2$ (R)  & 1.768 & 0.281   \\
\hline
    \end{tabular}
    \caption{Selected transitions with corresponding reference values for ${\Delta}E_{m, n}$ (in eV) and $f$ (in length gauge) obtained at the CC3/dAVQZ. V = valence, R = Rydberg.
    The asterisks highlight degenerate states (see main text).}
    \label{tab:Table-2}
\end{table}

While the reduction in MAE achieved by expanding the AVDZ basis set with respect to $\zeta$ is only marginal (MAEs for $f$ of {\MAEsmallsetCCTHAVDZ}, {\MAEsmallsetCCTHAVTZ}, and {\MAEsmallsetCCTHAVQZ} for AVDZ, AVTZ, and AVQZ, respectively), further augmenting the basis with diffuse functions brings a significant improvement (MAEs for $f$ of {\MAEsmallsetCCTHAVDZ}, {\MAEsmallsetCCTHdAVDZ}, and {\MAEsmallsetCCTHtAVDZ} for AVDZ, dAVDZ, and tAVDZ, respectively). The significant effect of additional diffuse functions stems from the fact that most of the transitions in this comparison include at least one state of Rydberg character. Using a triply-augmented basis set (tAVDZ or tAVTZ) changes the results only marginally and is, therefore, unnecessary. This further justifies the choice of a doubly-augmented basis set as a reference as opposed to a triply-augmented one.

Additionally, the differences between MAE and SDE, and MAE and RMSE, are much smaller in the case of doubly- (triply-)augmented basis sets hinting at a rather systematic error pattern.  

As stated above, the trends in $f$ and $\mu$ values are consistent. This should be the case since $f$ is proportional to $\mu$. However, unlike $\mu$, $f$ is influenced by the computed energy difference, which can introduce an additional source of error. Nevertheless, given the data in Fig.~\ref{fig:Figure-2}, we will focus solely on comparing $f$ below.

To further confirm the quality of QR-CC3/dAVTZ values, we investigate gauge invariance at this level of theory. It can be clearly seen in Table \ref{tab:Table-3} that the choice of gauge is insignificant as MAE, RMSE, and SDE are all lower than 0.005, and the MAPE is smaller than 0.1. Therefore, according to the gauge invariance metrics, the $f$ values obtained at this level of theory are very accurate.

Looking at the convergence of different gauges for all QR-CC approaches (see Table S40 in SI), we found three outcomes: i) the deviation of the velocity and mixed gauges from the length one is minimal at the QR-CC3 level; ii) the deviation of the mixed gauge from the length one decreases when one increases the level of theory, that is, QR-CC2 > QR-CCSD > QR-CC3; iii) the velocity gauge values obtained with QR-CCSD deviate from their length counterparts significantly more than in the case of QR-CC2. While investigating this last unexpected outcome, we found that, in a few cases, when the QR-CCSD velocity gauge differs significantly from the length one, the values of the velocity transition dipole moments of the left and right eigenvectors differ to a great extent. The reason behind this behavior remains unclear.

\begin{table}
\small
   \begin{tabular}{ l @{\hspace{8pt}} c @{\hspace{8pt}} c @{\hspace{8pt}} c @{\hspace{8pt}} c @{\hspace{8pt}} }  
   \hline
       gauge & MSE & MAE & RMSE & SDE \\
     \hline
velocity & 0.0018 & 0.0021 & 0.0043 & 0.0039  \\
mixed & 0.0009 & 0.0011 & 0.0021 & 0.0019  \\
\hline
    \end{tabular}
    \caption{MSE, MAE, RMSE, and SDE associated with oscillator strengths ($f$) obtained in the velocity and mixed gauges (CC3/dAVTZ). $f$ values obtained in the length gauge at the CC3/dAVTZ level serve as our reference.} 
    \label{tab:Table-3}
\end{table}

For additional verification, we extrapolated the oscillator strengths using a second-degree polynomial to obtain FCI estimates, as described in Sec.~\ref{sec:comp_det}. 
We examined the $1{\mathrm{A}}_2 \rightarrow 1{\mathrm{B}}_1$ transition
of water, the $2\Sigma^+ \rightarrow 3{\Sigma}^+$ and $1\Pi \rightarrow 2\Pi$
transitions of carbon monoxide, as well as the $1{\mathrm{A}}_2 \rightarrow 1{\mathrm{E}}$ and
$1{\mathrm{A}}_2 \rightarrow 2{\mathrm{A}}_1$ transitions of ammonia. The extrapolated FCI results within the dAVDZ basis
are summarized in Table \ref{tab:Table-4}, where they are compared with CC3 values. The CC3 
results exhibit an excellent agreement with extrapolated FCI values, with a
maximum deviation of 0.006 observed for the $1\Pi \rightarrow 2\Pi$ transition
of the carbon monoxide using the length gauge.

\begin{table}
\caption{Extrapolated FCI (exFCI) and CC3 oscillator strengths of the $1{\mathrm{A}}_2 \rightarrow 1{\mathrm{B}}_1$ 
transition of water, the $2\Sigma^+ \rightarrow 3{\Sigma}^+$ and $1\Pi \rightarrow 2\Pi$ transitions of carbon monoxide,
and the $1{\mathrm{A}}_2 \rightarrow 1{\mathrm{E}}$ and $1{\mathrm{A}}_2 \rightarrow 2{\mathrm{A}}_1$ transitions of ammonia,
calculated in the dAVDZ basis set.}
\label{tab:Table-4}
\small
\begin{tabular}{llccc}
\hline
Molecule \ \ & \ \ Transition \ \ & \ \ \ \ \ \ \ \ & \ \ \ \ exFCI \ \ \ \ & \ \ \ \ CC3 \ \ \ \ \\
\hline
Water    & $1{\mathrm{A}}_2 \rightarrow 1{\mathrm{B}}_1$ & $f^\text{l}$ & 0.2897(4) & 0.288 \\
         &                                               & $f^\text{v}$ & 0.2853(1) & 0.285 \\
         &                                               & $f^\text{m}$ & 0.2875(3) & 0.287 \\ 
CO       & $2\Sigma^+ \rightarrow 3{\Sigma}^+$           & $f^\text{l}$ & 0.0109(0) & 0.011 \\ 
         &                                               & $f^\text{v}$ & 0.0082(2) & 0.007 \\ 
         &                                               & $f^\text{m}$ & 0.0095(2) & 0.009 \\ 
         & $1\Pi \rightarrow 2\Pi$                       & $f^\text{l}$ & 0.1405(9) & 0.147 \\ 
         &                                               & $f^\text{v}$ & 0.152(2) \ \ & 0.151 \\
         &                                               & $f^\text{m}$ & 0.1466(5) & 0.149 \\
Ammonia  & $1{\mathrm{A}}_2 \rightarrow 1{\mathrm{E}}$   & $f^\text{l}$ & 0.3229(0) & 0.323 \\
         &                                               & $f^\text{v}$ & 0.3223(0) & 0.323 \\
         &                                               & $f^\text{m}$ & 0.3226(0) & 0.323 \\
         & $1{\mathrm{A}}_2 \rightarrow 2{\mathrm{A}}_1$ & $f^\text{l}$ & 0.2972(0) & 0.298 \\
         &                                               & $f^\text{v}$ & 0.2968(1) & 0.298 \\
         &                                               & $f^\text{m}$ & 0.2970(0) & 0.298 \\
\hline
\end{tabular}
\end{table}

\begin{table*}[tp]
\small
\centering
\caption{Full list of CC3/dAVTZ  transitions from state $n$ to state $m$. Transitions energies ${\Delta}E_{0,n}$, ${\Delta}E_{0,m}$, and ${\Delta}E_{m,n}$ (in eV), the percentage of single excitation character (${\%}T_{1}$) of the two states and ESA oscillator strengths in length, velocity and mixed gauges, $f^{\mathrm{l}}$, $f^{\mathrm{v}}$ and $f^{\mathrm{m}}$, respectively. Rydberg = R, valence = V. The asterisks highlight degenerate states (see main text).}
\vspace{-0.3 cm}
\label{tab:Table-5}
\begin{tabular}{%
>{\raggedright\arraybackslash}p{3.5cm}  
>{\raggedright\arraybackslash}p{1.8cm} 
>{\centering\arraybackslash}p{1cm} 
>{\centering\arraybackslash}p{1cm} 
>{\raggedright\arraybackslash}p{1.8cm} 
>{\centering\arraybackslash}p{1cm} 
>{\centering\arraybackslash}p{1cm} 
>{\centering\arraybackslash}p{1cm} 
>{\centering\arraybackslash}p{1cm} 
>{\centering\arraybackslash}p{1cm} 
>{\centering\arraybackslash}p{1cm} }
\hline
 & \multicolumn{3}{c}{Initial state $n$} & \multicolumn{3}{c}{Final state $m$}& \multicolumn{4}{c}{Transition}\\ 
 \cline{2-4} \cline{5-7} \cline{8-11}
Molecule & state & ${\Delta}E_{0,n}$ & ${\%}T_{1}$ & state  & ${\Delta}E_{0,m}$  &  ${\%}T_{1}$ & ${\Delta}E_{{m,n}}$ & $f^{\mathrm{l}}$ & $f^{\mathrm{v}}$ & $f^{\mathrm{m}}$ \\ 
\hline
Acetone & 1$\mathrm{B}_2$ (R)  & 6.418 & 90.6 & 2$\mathrm{A}_1$ (R)  & 7.450& 90.6 & 1.032 & 0.256 & 0.259 & 0.257   \\
 & 1$\mathrm{B}_2$ (R)  &  6.418 & 91.1 & 2$\mathrm{B}_2$ (R)   & 7.456 & 91.1 & 1.037 & 0.267 &  0.268 & 0.268  \\
 & 1$\mathrm{B}_2$ (R)  &  6.418 & 91.0 & 2$\mathrm{A}_2$ (R)   & 7.382 & 91.0 & 0.964 & 0.361 &  0.364 & 0.363  \\
Acrolein & 1A" (V)  & 3.741 & 79.5 & 2A" (V)   & 6.743 & 79.5 & 3.002 & 0.036 & 0.039 & 0.037   \\
Ammonia & 1$\mathrm{A}_2$ (R)  & 6.572 & 93.7 & 1E (R)*  & 8.117 & 93.7 & 1.545 & 0.320 & 0.321 & 0.321   \\
 & 1$\mathrm{A}_2$ (R)  &  6.572 & 93.4 & 2$\mathrm{A}_1$ (R)   & 8.564 & 93.4 & 1.992 & 0.295 &  0.296 & 0.296  \\
 & 1E (R)* &  8.117 & 93.6 & 2$\mathrm{A}_2$ (R)   & 9.121 & 93.6 & 1.003 & 0.139 &  0.143 & 0.141  \\
 & 2$\mathrm{A}_1$ (R)  &  8.564 & 93.6 & 2$\mathrm{A}_2$ (R)   & 9.121 & 93.6 & 0.556 & 0.252 &  0.278 & 0.264  \\
Benzene & 1$\mathrm{E}_\mathrm{1g}$ (R)* & 6.504 & 92.8 & 1$\mathrm{A}_\mathrm{2u}$ (R)  & 7.053& 92.8 & 0.549 & 0.132 & 0.133 & 0.133   \\
 & 1$\mathrm{E}_\mathrm{1g}$ (R)* &  6.504 & 92.9 & 1$\mathrm{E}_\mathrm{2u}$ (R)*  & 7.117 & 92.9 & 0.613 & 0.138 &  0.140 & 0.139  \\
Butadiene & 1$\mathrm{B}_\mathrm{g}$ (R)  & 6.310 & 94.2 & 1$\mathrm{A}_\mathrm{u}$ (R)  & 6.626& 94.2 & 0.316 & 0.122 & 0.124 & 0.124   \\
 & 1$\mathrm{B}_\mathrm{g}$ (R)  & 6.310 & 94.2 & 2$\mathrm{A}_\mathrm{u}$ (R)   & 6.780 & 94.2 & 0.470 & 0.165 &  0.167 & 0.166  \\
 & 1$\mathrm{A}_\mathrm{u}$ (R)  &  6.626 & 94.4 & 2$\mathrm{B}_\mathrm{g}$ (R)   & 7.447 & 94.4 & 0.821 & 0.062 &  0.062 & 0.062  \\
 & 2$\mathrm{A}_\mathrm{u}$ (R)  &  6.780 & 94.4 & 2$\mathrm{B}_\mathrm{g}$ (R)   & 7.447 & 94.4 & 0.667 & 0.367 &  0.369 & 0.368  \\
Carbon monoxide & 2$\mathrm{\Sigma}^{+}$ (R)  & 10.676 & 92.4 & 3$\mathrm{\Sigma}^{+}$ (R)   & 11.285 & 92.4 & 0.609 & 0.146 & 0.146 & 0.147   \\
 & 2$\mathrm{\Sigma}^{+}$ (R)  &  10.676 & 92.4 & 2$\mathrm{\Pi}$ (R)*  & 11.414 & 92.4 & 0.738 & 0.256 &  0.261 & 0.258  \\
 & 2$\mathrm{\Pi}$ (R)* & 11.414 & 92.4 & 1$\mathrm{\Pi}$ (V)*  & 8.481 & 92.4 & 2.933 & 0.010 &  0.009 & 0.009  \\
 & 2$\mathrm{\Sigma}^{+}$ (R)  &  10.676 & 93.1 & 1$\mathrm{\Pi}$ (V)*  & 8.481 & 93.1 & 2.195 & 0.025 &  0.025 & 0.025  \\
 & 3$\mathrm{\Sigma}^{+}$ (R)  &  11.285 & 93.1 & 1$\mathrm{\Pi}$ (V)*  & 8.481 & 93.1 & 2.804 & 0.010 &  0.010 & 0.010  \\
Cyclopentadiene & 1$\mathrm{B}_2$ (V)  & 5.538 & 93.8 & 2$\mathrm{A}_1$ (V)  & 6.567& 93.8 & 1.029 & 0.023 & 0.032 & 0.027   \\
 & 1$\mathrm{A}_2$ (R)  & 5.759 & 94.0 & 1$\mathrm{B}_1$ (R)   & 6.375 & 94.0 & 0.616 & 0.235 &  0.236 & 0.236  \\
 & 1$\mathrm{A}_2$ (R)  & 5.759 & 94.0 & 2$\mathrm{B}_2$ (R)   & 6.448 & 94.0 & 0.689 & 0.282 &  0.284 & 0.283  \\
 & 1$\mathrm{A}_2$ (R)  &  5.759 & 93.9 & 2$\mathrm{A}_2$ (R)   & 6.412 & 93.9 & 0.653 & 0.252 &  0.254 & 0.253  \\
Cyclopropene & 2$\mathrm{B}_1$ (R)  & 6.921 & 95.1 & 3$\mathrm{B}_1$ (R)   & 7.294 & 95.1 & 0.373 & 0.103 & 0.102 & 0.103   \\
Diazirine & 1$\mathrm{B}_1$ (V)  & 4.113 & 92.5 & 2$\mathrm{A}_1$ (V)  & 7.972& 92.5 & 3.859 & 0.014 & 0.015 & 0.014   \\
 & 1$\mathrm{B}_2$ (R)  & 7.442 & 93.5 & 2$\mathrm{A}_1$ (V)   & 7.972 & 93.5 & 0.530 & 0.085 &  0.084 & 0.084  \\
Diazomethane & 1$\mathrm{A}_2$ (V)  & 3.071 & 90.1 & 1$\mathrm{B}_1$ (R)  & 5.436& 90.1 & 2.364 & 0.012 & 0.012 & 0.012   \\
Formaldehyde & 1$\mathrm{B}_2$ (R)  & 7.173 & 91.7 & 2$\mathrm{A}_1$ (R)  & 8.144& 91.7 & 0.972 & 0.297 & 0.301 & 0.299   \\
 & 1$\mathrm{B}_2$ (R)  &  7.173 & 92.0 & 2$\mathrm{A}_2$ (R)   & 8.386 & 92.0 & 1.214 & 0.278 &  0.279 & 0.279  \\
 & 1$\mathrm{A}_2$ (V)  & 3.968 & 91.5 & 1$\mathrm{B}_2$ (R)   & 7.173 & 91.5 & 3.205 & 0.019 &  0.020 & 0.020  \\
 & 2$\mathrm{B}_2$ (R)  &  7.995 & 92.0 & 2$\mathrm{A}_2$ (R)   & 8.386 & 92.0 & 0.392 & 0.011 &  0.011 & 0.011  \\
 & 1$\mathrm{A}_2$ (V)  & 3.968 & 91.5 & 2$\mathrm{A}_2$ (R)   & 8.386 & 91.5 & 4.419 & 0.015 &  0.014 & 0.014  \\
Furan & 1$\mathrm{A}_2$ (R)  & 6.061 & 93.8 & 2$\mathrm{B}_2$ (R)  & 6.883& 93.8 & 0.822 & 0.271 & 0.272 & 0.271   \\
 & 1$\mathrm{A}_2$ (R)  & 6.061 & 93.8 & 1$\mathrm{B}_1$ (R)   & 6.606 & 93.8 & 0.544 & 0.220 &  0.220 & 0.220  \\
 & 1$\mathrm{A}_2$ (R)  & 6.061 & 93.8 & 2$\mathrm{A}_2$ (R)   & 6.754 & 93.8 & 0.693 & 0.268 &  0.269 & 0.269  \\
Glyoxal & 1$\mathrm{A}_\mathrm{u}$ (V)  & 2.874 & 88.3 & 1$\mathrm{B}_\mathrm{g}$ (V)   & 4.267 & 88.3 & 1.393 & 0.020 & 0.021 & 0.021   \\
 & 1$\mathrm{A}_\mathrm{u}$ (V)  &  2.874 & 84.0 & 2$\mathrm{B}_\mathrm{g}$ (V)   & 6.566 & 84.0 & 3.692 & 0.113 &  0.117 & 0.115  \\
Ketene & 1$\mathrm{B}_1$ (R)  & 5.948 & 94.3 & 2$\mathrm{A}_2$ (R)   & 7.109 & 94.3 & 1.161 & 0.336 & 0.338 & 0.337   \\
 & 1$\mathrm{B}_1$ (R)  & 5.948 & 93.9 & 2$\mathrm{A}_1$ (V)   & 7.085 & 93.9 & 1.136 & 0.176 &  0.174 & 0.175  \\
Mcp & 1$\mathrm{B}_1$ (R)  & 5.422 & 93.3 & 1$\mathrm{A}_2$ (R)   & 5.928 & 93.3 & 0.506 & 0.188 & 0.189 & 0.189   \\
 & 1$\mathrm{B}_2$ (V)  &  4.311 & 93.3 & 1$\mathrm{A}_2$ (R)   & 5.928 & 93.3 & 1.616 & 0.023 &  0.025 & 0.024  \\
Pyrazine & 2$\mathrm{A}_\mathrm{g}$ (R)  & 6.638 & 91.5 & 2$\mathrm{B}_\mathrm{1u}$ (R)   & 7.409 & 91.5 & 0.771 & 0.198 & 0.197 & 0.198   \\
 & 2$\mathrm{A}_\mathrm{g}$ (R)  &  6.638 & 90.8 & 2$\mathrm{B}_\mathrm{2u}$ (R)   & 7.229 & 90.8 & 0.591 & 0.240 &  0.244 & 0.242  \\
 & 1$\mathrm{B}_\mathrm{1u}$ (V)  &  7.409 & 91.2 & 1$\mathrm{B}_\mathrm{3g}$ (R)   & 7.951 & 91.2 & 1.090 & 0.004 &  0.004 & 0.004  \\
 & 2$\mathrm{B}_\mathrm{2u}$ (R)  &  7.229 & 91.2 & 1$\mathrm{B}_\mathrm{3g}$ (R)   & 7.951 & 91.2 & 0.722 & 0.201 &  0.201 & 0.201  \\
Pyridazine & 1$\mathrm{A}_2$ (V)  & 4.372 & 86.9 & 2$\mathrm{B}_1$ (V)  & 6.366& 86.9 & 1.995 & 0.012 & 0.010 & 0.011   \\
Pyrrole & 1$\mathrm{A}_2$ (R)  & 5.233 & 92.9 & 1$\mathrm{B}_1$ (R)  & 5.967& 92.9 & 0.734 & 0.091 & 0.091 & 0.091   \\
 & 1$\mathrm{A}_2$ (R)  &  5.233 & 93.0 & 2$\mathrm{A}_2$ (R)   & 5.980 & 93.0 & 0.747 & 0.171 &  0.171 & 0.171  \\
Silylidene & 1$\mathrm{A}_2$ (V)  & 2.155 & 88.0 & 1$\mathrm{B}_2$ (V)  & 3.780& 88.0 & 1.624 & 0.003 & 0.002 & 0.003   \\
Tetrazine & 1$\mathrm{B}_\mathrm{3u}$ (V)  & 2.454 & 83.1 & 1$\mathrm{B}_\mathrm{1g}$ (V)   & 4.910 & 83.1 & 2.456 & 0.061 & 0.068 & 0.065   \\
 & 1$\mathrm{B}_\mathrm{3u}$ (V)  &  2.454 & 81.8 & 1$\mathrm{B}_\mathrm{2g}$ (V)   & 5.463 & 81.8 & 3.009 & 0.017 &  0.017 & 0.017  \\
Thiophene & 1$\mathrm{A}_2$ (R)  & 6.121 & 92.7 & 1$\mathrm{B}_1$ (R)  & 6.131& 92.7 & 0.009 & 0.001 & -0.004 & 0.000   \\
Water & 1$\mathrm{B}_1$ (R)  & 7.600 & 93.5 & 1$\mathrm{A}_2$ (R)   & 9.368 & 93.5 & 1.768 & 0.283 & 0.284 & 0.283   \\
\hline
\end{tabular}
\end{table*}

\subsection{Basis set effects}

\begin{figure}[htp]
   \centering
	    \includegraphics[width=.8\linewidth]{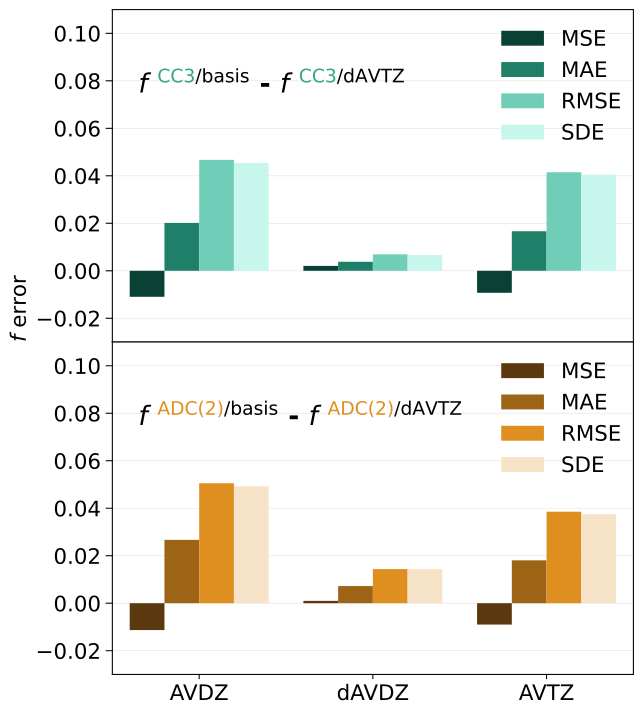}
   \caption{MSE, MAE, RMSE, and SDE of $f$, in length gauge, computed with three different basis sets (AVDZ, dAVDZ, and AVTZ). Top: CC3 results compared with CC3/dAVTZ; bottom: ADC(2) results compared with ADC(2)/dAVTZ.}
 	   \label{fig:Figure-3}
    \end{figure}

After establishing that the dAVTZ basis set is sufficiently large for computing oscillator strengths, we explore the effects of using smaller basis sets. We examine the $f$ values obtained with AVDZ, dAVDZ, and AVTZ at the QR-CC3, QR-CCSD, QR-CC2, and ISR-ADC(2) levels of theory, comparing them to the corresponding dAVTZ values. By excluding the computationally demanding triple-augmented and quadruple-$\zeta$ basis sets, we were able to include in this comparison all {\alltrans} transitions, listed in Table \ref{tab:Table-5}. QR-CC3 and ISR-ADC(2) statistical values obtained with different basis sets are shown in Fig.~\ref{fig:Figure-3}. Additional results can be found in the SI (Tables S41, S42). 

Upon looking at the QR-CC3 results, one can see that the trends are consistent with the findings obtained for the smaller set of transitions (see above). Augmentation with extra diffuse functions is more crucial than increasing $\zeta$-multiplicity. Indeed, the AVTZ results (MAE = {\MAEallCCTHAVTZ}, RMSE = {\RMSEallCCTHAVTZ}, and SDE = {\STEallCCTHAVTZ}) do not significantly differ from their AVDZ counterparts (MSE = {\MAEallCCTHAVDZ}, RMSE = {\RMSEallCCTHAVDZ}, and SDE = {\STEallCCTHAVDZ}). On the other hand, the dAVDZ results (MAE = {\MAEallCCTHdAVDZ}, RMSE = {\RMSEallCCTHdAVDZ}, and SDE = {\STEallCCTHdAVDZ}) are much closer to the reference values than their singly-augmented AVDZ equivalents.

\begin{figure}[htp]
   \centering
	    \includegraphics[width=.8\linewidth]{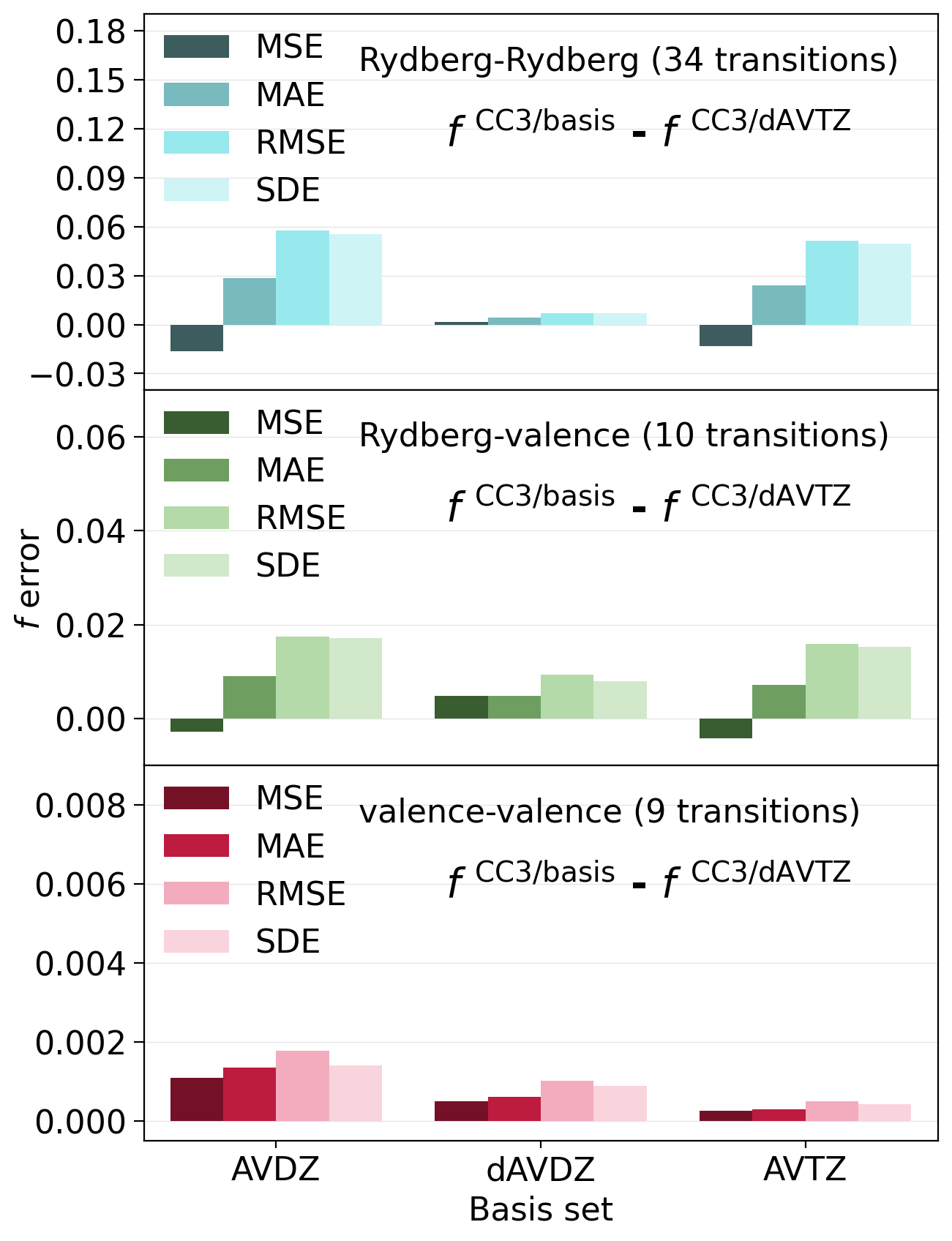}
   \caption{MSE, MAE, RMSE, and SDE of $f$, in length gauge, computed with different basis sets at the CC3 level of theory for various sets of transitions: Rydberg-Rydberg (top), Rydberg-valence (center), and valence-valence (bottom). The reference values are obtained at the CC3/dAVTZ level. Note the difference in the scales of the vertical axis.}
	   \label{fig:Figure-4}
    \end{figure}

We further divide the QR-CC3 results based on the nature of states involved in the transition  (Fig.~\ref{fig:Figure-4}): Rydberg-Rydberg, Rydberg-valence, and valence-valence. For the first category of transitions, the trend remains consistent with the pattern observed when all transitions are considered together, which is expected as the majority of transitions, {\rydrydtrans} out of {\alltrans}, are of Rydberg-Rydberg nature. Examining the Rydberg-valence and valence-valence transitions reveals that the importance of additional diffuse functions diminishes, as anticipated, with a clear preference for higher $\zeta$-multiplicity over extra diffuse functions in the case of ESA between two valence states. Additionally, the errors of QR-CC3 decrease across the transition types, following the series Rydberg-Rydberg $\rightarrow$ Rydberg-valence $\rightarrow$ valence-valence. This error reduction is more pronounced with the singly augmented basis sets, whereas only a slight improvement is observed with dAVDZ. Overall, the MAE in $f$ at the QR-CC3 level remains consistently small across different basis sets, not exceeding 0.03.

We also investigate basis set effects at lower levels of theory by comparing $f$ values computed with AVDZ, dAVDZ, and AVTZ at the QR-CCSD, QR-CC2, and ISR-ADC(2) levels, against their corresponding dAVTZ values obtained at the same level of theory. We show here only the ISR-ADC(2) example in Fig.~{\ref{fig:Figure-3}. In all cases, we found that including additional diffuse functions is more crucial than increasing $\zeta$-multiplicity, which is consistent with our findings at the QR-CC3 level of theory (see above). The trend remains consistent when comparing the ISR-ADC(2) values obtained with different basis sets to our reference CC3/dAVTZ data. The MAEs diminish across ISR-ADC(2)/AVDZ (MAE = {\MAEallADCTWCCTHAVDZ}), ISR-ADC(2)/AVTZ (MAE = {\MAEallADCTWCCTHAVTZ}), ISR-ADC(2)/dAVDZ (MAE = {\MAEallADCTWCCTHdAVDZ}), ISR-ADC(2)/dAVTZ (MAE = {\MAEallADCTWCCTHdAVTZ}), as excpected.

At this stage, we would like to mention that the QR-CC3/dAVTZ calculations are extremely computationally demanding. This issue could be alleviated by employing the dAVDZ basis set instead of dAVTZ, as the errors at the QR-CC3 level remain relatively small. Indeed, in our full testing set of {\alltrans} transitions, the largest deviation of $f$ between results obtained with dAVDZ and dAVTZ occurs for the $1\mathrm{A}_2$ (Rydberg, $\pi$}-3s) $\rightarrow$ $1\mathrm{B}_1$ (Rydberg, {$\pi$}-3p) transition of thiophene where the QR-CC3/dAVTZ (dAVDZ) $f$ is 0.001 (0.023). Clearly, it is not a chemically significant deviation as both values of $f$ would characterize the transition as a low-intensity one. Nevertheless, we note that this deviation is not due to state mixing caused by basis set expansion as an examination of the orbital compositions for both states with both basis sets revealed no substantial state mixing. We opted to stick with the larger dAVTZ basis for the remainder of this study.

  \begin{figure*}[ht]
   \centering
	    \includegraphics[width=.7\linewidth]{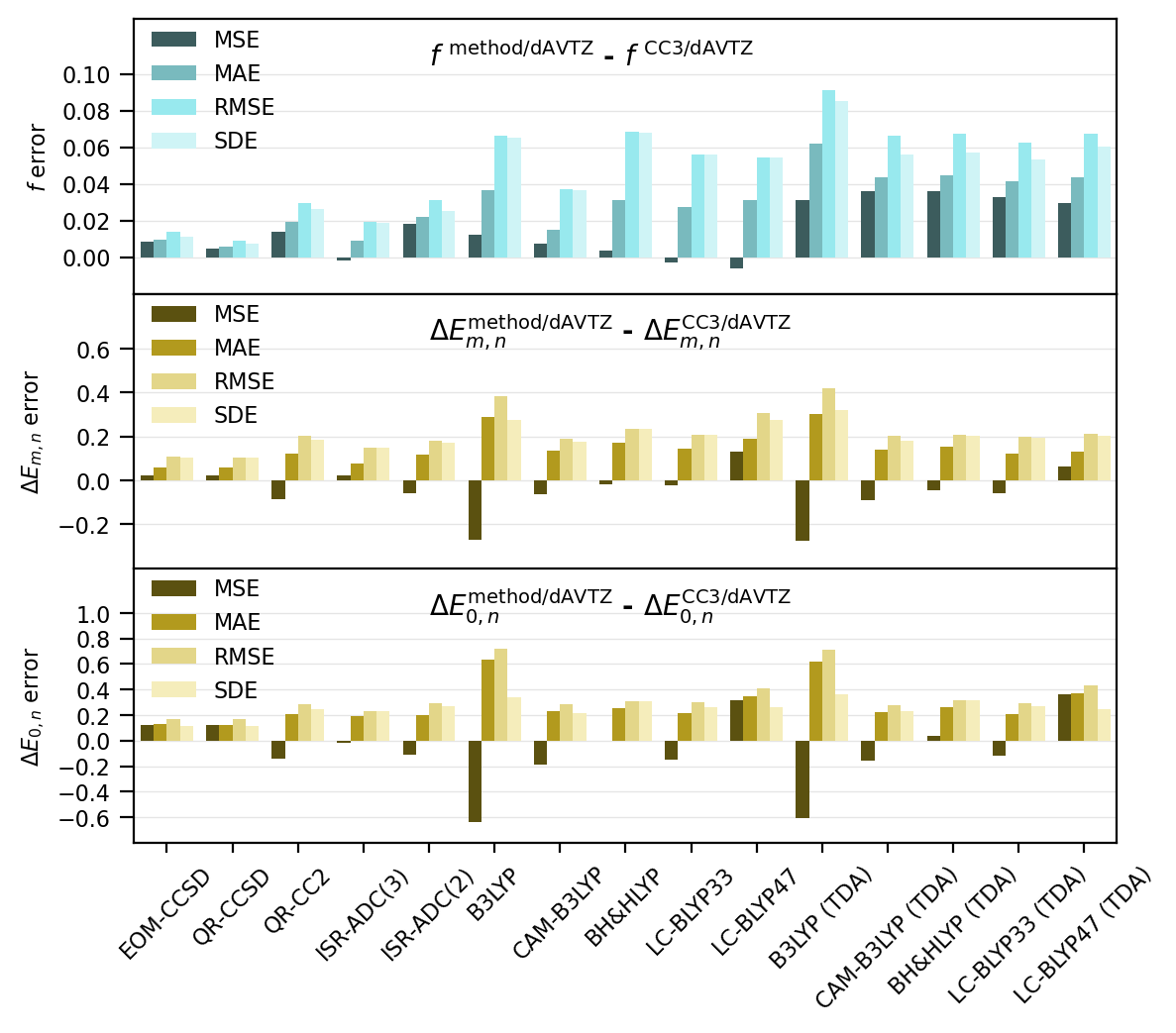}
   \caption{Comparison of MSE, MAE, RMSE, and SDE for all methods. Top: $f$, in length gauge; center: ${\Delta}E_{m,n}$ in eV; bottom: ${\Delta}E_{0,n}$ in eV. ${\Delta}E_{m,n}$ and $f$ values are compared with QR-CC3/dAVTZ and ${\Delta}E_{0,n}$ with CC3/dAVTZ.}
            \label{fig:Figure-5}
    \end{figure*}

\subsection{Methods benchmark}

We compare the errors in $f$ (length gauge), ${\Delta}E_{0,n}$, and ${\Delta}E_{m, n}$ obtained with all methods (dAVTZ basis set) against QR-CC3/dAVTZ in Fig.~\ref{fig:Figure-5}. The ${\Delta}E_{{0,n}}$values correspond to the GSA vertical transition energies ({\allstates} excited states) and ${\Delta}E_{m, n} = {\Delta}E_{0, n} - {\Delta}E_{0, m}$ corresponds to the difference of the two GSA values involved in the corresponding ESA transition ({\alltrans} transitions). The MAEs of $f$ for the wavefunction methods grow in the series: QR-CCSD ({\MAEallCCSD}), ISR-ADC(3) ({\MAEallADCTH}), EOM-CCSD ({\MAEallEOMCCSD}), QR-CC2 ({\MAEallCCTW}), and ISR-ADC(2) ({\MAEallADCTW}). We see that QR-CCSD is the closest to our reference, with EOM-CCSD and ISR-ADC(3) only slightly behind. We would like to stress the excellent performance of ISR-ADC(3) which yields results comparable to EOM-CCSD, consistent with previous findings for GSA by some of us. \cite{sarkar_benchmarking_2021} While ADC(3) is not always trustworthy for transition energies from the ground state (see below), it can provide accurate properties. When we compare the MAE in ${\Delta}E_{m, n}$ with the MAE in ${\Delta}E_{{0,n}}$, we see that the errors are smaller for ${\Delta}E_{m, n}$ than for ${\Delta}E_{{0,n}}$, which is logical since the GSA energy differences are often larger than the ESA ones. 
This is beneficial for calculating ESA $f$ values, as they are computed from $\mu$ and ${\Delta}E_{m, n}$ rather than ${\Delta}E_{{0,n}}$, as in the case of GSA $f$ values. The difference between the MAEs of ${\Delta}E_{{0,n}}$ and ${\Delta}E_{m, n}$ for the wavefunction approaches grows across the following series: CCSD ({\MAEallCCSDEdif} eV),  CC2 ({\MAEallCCTWEdif} eV), ADC(2) ({\MAEallADCTWEdif} eV), and  ADC(3) ({\MAEallADCTHEdif} eV). At this stage, it is worth reminding that the energies obtained with QR-CCSD and EOM-CCSD are identical within numerical error. The RMSE and SDE in $f$ obtained from the wavefunction approaches are very close to the MAE values, with ISR-ADC(3) and second-order approaches [QR-CC2 and ISR-ADC(2)] giving larger deviations between these statistical indicators. This is an important feature of these methods indicating a rather systematic error behavior, tight error distributions, and low frequency of outliers. This is in contrast with QR-TDDFT, as discussed below. 

  \begin{figure*}[ht]
   \centering
	    \includegraphics[width=.7\linewidth]{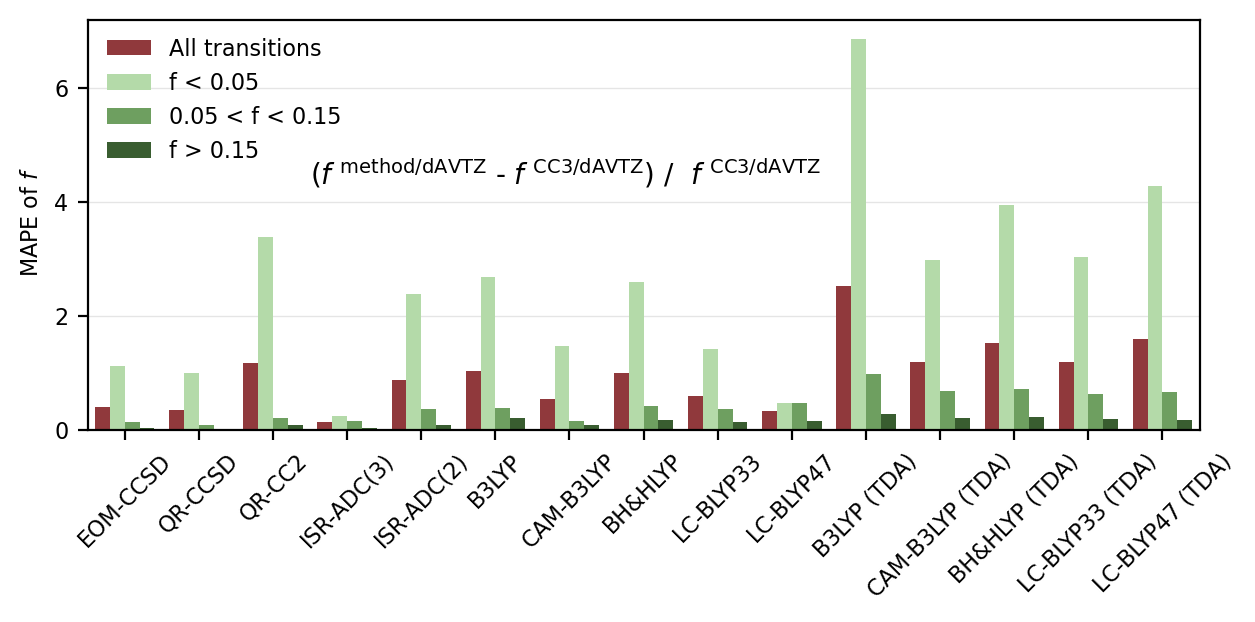}
   \caption{Comparison of MAPE  in $f$ values of all methods divided by the magnitude of the computed $f$ into three groups: $f < 0.05$, $0.05 < f < 0.15$, and $f > 0.15$. The reference values are obtained at the CC3/dAVTZ level in length gauge}
            \label{fig:Figure-6}
    \end{figure*}

Upon analyzing the QR-TDDFT results, two general features become apparent. First, the difference between the RMSE and MAE values of $f$ is larger than those observed with wavefunction approaches. The same holds for the differences between SDE and MAE. As mentioned earlier, this behavior is in stark contrast with wavefunction approaches and has been reported by some of us previously for two-photon absorption cross-sections computed with QR-TDDFT. \cite{naim_two-photon_2024} This highlights a key characteristic of QR-TDDFT: it can produce highly accurate transition properties, yet it may also yield significantly inaccurate results.
The second feature is the impact of the TDA on the calculated $f$ values and transition energies. While the TDA offers a reduction in computational cost, this comes at the expense of a noticeable deterioration in the computed $f$ values, as reflected by an increase in the MAE. The magnitude of this effect depends on the functional. Nevertheless, the ranking of functionals, based on the MAE of $f$, remains relatively consistent between full TD-DFT and TDA calculations.
For full TD-DFT, the ranking is as follows: CAM-B3LYP ({\MAEallTDDFTCAMBTHLYP}), LC-BLYP33 ({\MAEallTDDFTLCBLYPTHTH}), LC-BLYP47 ({\MAEallTDDFTLCBLYPFOSE}), BH\&HLYP ({\MAEallTDDFTBHandHLYP}), and B3LYP ({\MAEallTDDFTBTHLYP}).
For TDA, the ranking becomes: LC-BLYP33 ({\MAEallTDALCBLYPTHTH}), CAM-B3LYP ({\MAEallTDACAMBTHLYP}), LC-BLYP47 ({\MAEallTDALCBLYPFOSE}), BH\&HLYP ({\MAEallTDABHandHLYP}), and B3LYP ({\MAEallTDABTHLYP}).

Interestingly, the effect of the TDA originates, in practice, solely from the $\mu$ values as the differences between the energies obtained with and without TDA are negligible for ESA. Similarly to what we found for wavefunction approaches, the cancellation of errors introduced by the subtraction of ${\Delta}E_{{0,n}}$ indeed improves the ${\Delta}E_{m, n}$ values. The difference between the MAEs of ${\Delta}E_{{0,n}}$ and the MAEs of ${\Delta}E_{m, n}$ of the TD-DFT approaches grows in the following series (see Table S45 in the SI):
LC-BLYP33 ({\MAEallTDDFTLCBLYPTHTHEdif} eV),
TDA-CAM-B3LYP ({\MAEallTDACAMBTHLYPEdif} eV),
BH\&HLYP ({\MAEallTDDFTBHandHLYPEdif} eV),
TDA-LC-BLYP33 ({\MAEallTDALCBLYPTHTHEdif} eV),
CAM-B3LYP ({\MAEallTDDFTCAMBTHLYPEdif} eV),
TDA-BH\&HLYP ({\MAEallTDABHandHLYPEdif} eV),
LC-BLYP47 ({\MAEallTDDFTLCBLYPFOSEEdif} eV),
TDA-LC-BLYP47 ({\MAEallTDALCBLYPFOSEEdif} eV),
TDA-B3LYP ({\MAEallTDABTHLYPEdif} eV),
and B3LYP ({\MAEallTDDFTBTHLYPEdif} eV).
While B3LYP largely undershoots the GSA energies in the present set, the errors become reasonable for the ESA energies.

From the analysis of all transitions, it is clear that CAM-B3LYP, without TDA, appears as the XCF of choice for computing ESA $f$ values at the TD-DFT level, with the lowest MAE ({\MAEallTDDFTCAMBTHLYP}), RMSE ({\RMSEallTDDFTCAMBTHLYP}) and SDE ({\STEallTDDFTCAMBTHLYP}) from all the tested functionals. The second best for ESA ($f$) is likely LC-BLYP33 with a MAE of {\MAEallTDDFTLCBLYPTHTH}, a RMSE of {\RMSEallTDDFTLCBLYPTHTH}, and a SDE of {\STEallTDDFTLCBLYPTHTH}. LC-BLYP47 provides rather similar errors (see the SI Table S44). Quite interestingly, CAM-B3LYP even outperforms both QR-CC2 and ISR-ADC(2) in terms of MAE but not for the SDE.

In short, it appears that QR-TDDFT may deliver average errors similar to, or even smaller than, those of second-order wavefunction methods when an adequate functional is selected. However, the spread of the errors and the number of outliers remain larger with QR-TDDFT.

When one considers the MAPEs of $f$ values instead of MAEs (see Fig.~\ref{fig:Figure-6}), one finds that ISR-ADC(3) and LC-BLYP47 (without TDA) perform unexpectedly well. Specifically, ISR-ADC(3) (MAPE = {\MAPEallADCTH}) outperforms all other methods while LC-BLYP47 (MAPE = {\MAPEallTDDFTLCBLYPFOSE}) ranks highest among the tested functionals. For the other methods, no significant deviations from the trends observed with MAEs were found. To further examine this outcome, we divided all transitions into three groups based on the magnitude of the computed $f$ values: transitions with $f<0.05$, transitions with $0.05<f<0.15$, and transitions with $f>0.15$ (see Fig.~\ref{fig:Figure-6}). While most methods show decreasing percentage errors as $f$ increases, most notably for $f < 0.05$, this trend does not hold for ISR-ADC(3) and LC-BLYP47. For these two models, the MAPE remains nearly constant when transitioning from moderately bright transitions ($0.05 < f < 0.15$) to dark transitions ($f < 0.05$). This shows that ISR-ADC(3) and LC-BLYP47 perform both very well for transitions with (very) small $f$ values, compared with the other tested methods. The reason behind this unexpected outcome remains unclear to us. Since real-life applications are mostly interested in bright ESA transitions with large $f$ values, it is likely of limited practical interest. 

  \begin{figure*}[ht]
   \centering
	    \includegraphics[width=.8\linewidth]{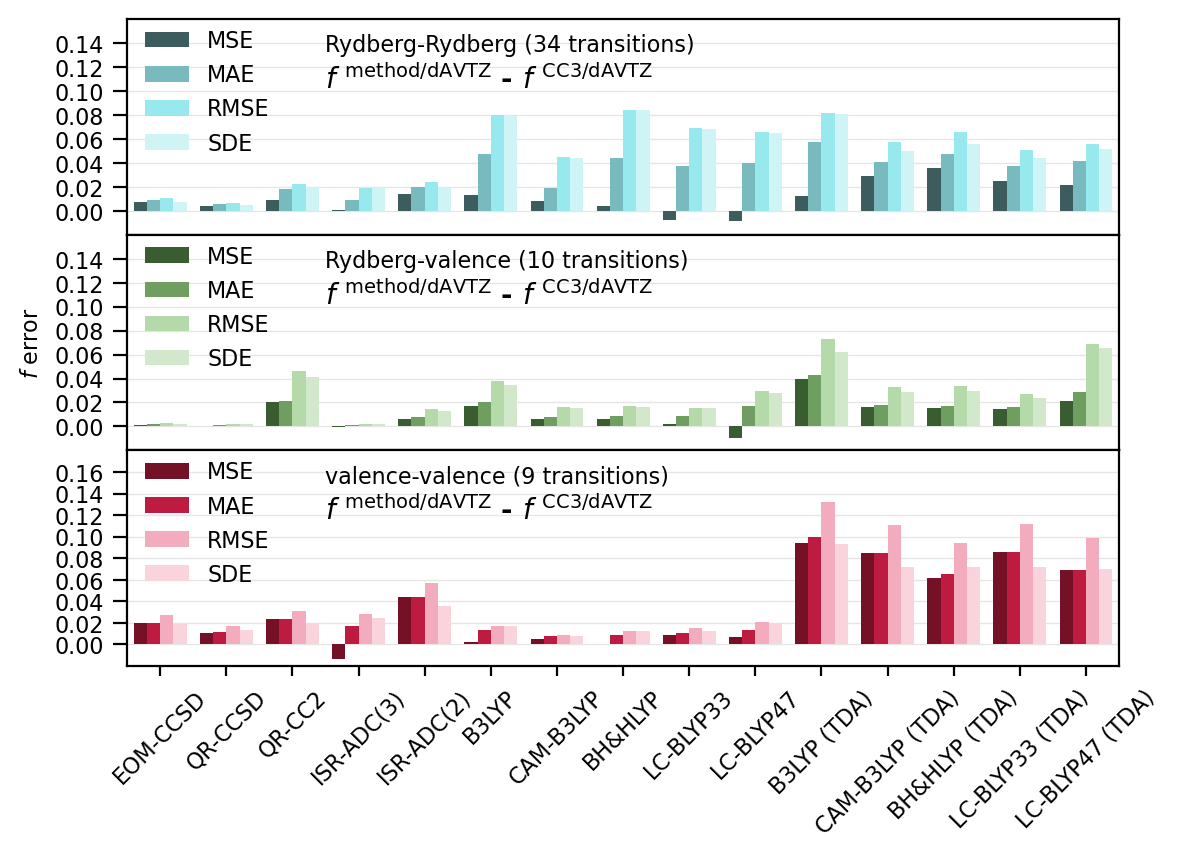}
   \caption{Comparison of errors in $f$ for all methods divided by the character of the states involved. Top: Rydberg-Rydberg; center: Rydberg-valence; bottom: valence-valence. See the caption of Fig.~\ref{fig:Figure-5} for more details.}
            \label{fig:Figure-7}
    \end{figure*}

To further characterize the performance of all methods, we divided the transitions by the character of states involved: Rydberg-Rydberg, Rydberg-valence, and valence-valence; as well as by the molecular size: molecules containing 1 to 3 non-hydrogen atoms, and molecules containing 4 to 6 non-hydrogen atoms (see Fig.~\ref{fig:Figure-1}). 

When looking at the influence of the nature of the states on $f$ values (Fig.~\ref{fig:Figure-7}), one observes that the performance of QR-TDDFT (without TDA) improves as valence states are included in the ESA transition. The RMSE and SDE become comparable with the MAE. This is expected, as firstly, the valence states are generally easier to model, and secondly, the amount of transitions involving valence states is lower, {\rydvaltrans} Rydberg-valence and {\valvaltrans} valence-valence compared with {\rydrydtrans} Rydberg-Rydberg excitations, making it less likely for strong outliers to occur. Unexpectedly, we do not observe the same trend when the TDA is enforced, as the performance is slightly worse for valence-valence transitions than for the Rydberg-Rydberg ones, the lowest deviation being observed for the Rydberg-valence transitions. The wavefunction methods show the same trends. In fact, full TD-DFT outperforms most of the wavefunction methods and is comparable to QR-CCSD when only valence-valence transitions are considered. Interestingly, BH\&HLYP and B3LYP perform similarly for Rydberg-Rydberg transitions although one would expect BH\&HLYP, a global hybrid with 50{\%} of exact exchange, to perform much better for Rydberg states than B3LYP, a global hybrid with 20{\%} of exact exchange. It is well-known that global hybrids with a low percentage of exact exchange tend to perform poorly at larger interelectronic distances due to the wrong behavior of the exchange-correlation kernel.\cite{tozer_improving_1998, casida_molecular_1998} Nevertheless, most states here are Rydberg-Rydberg in nature. In practice, the interesting, bright ESA transitions often involve states with significant Rydberg character. Indeed, in our set, only one valence-valence transition exhibits $f$ higher than 0.09 ($1{\mathrm{B}}_{\mathrm{g}} {\rightarrow} 1{\mathrm{A}}_{\mathrm{u}}$ transition of glyoxal).  

  \begin{figure*}[ht]
   \centering
	    \includegraphics[width=.8\linewidth]{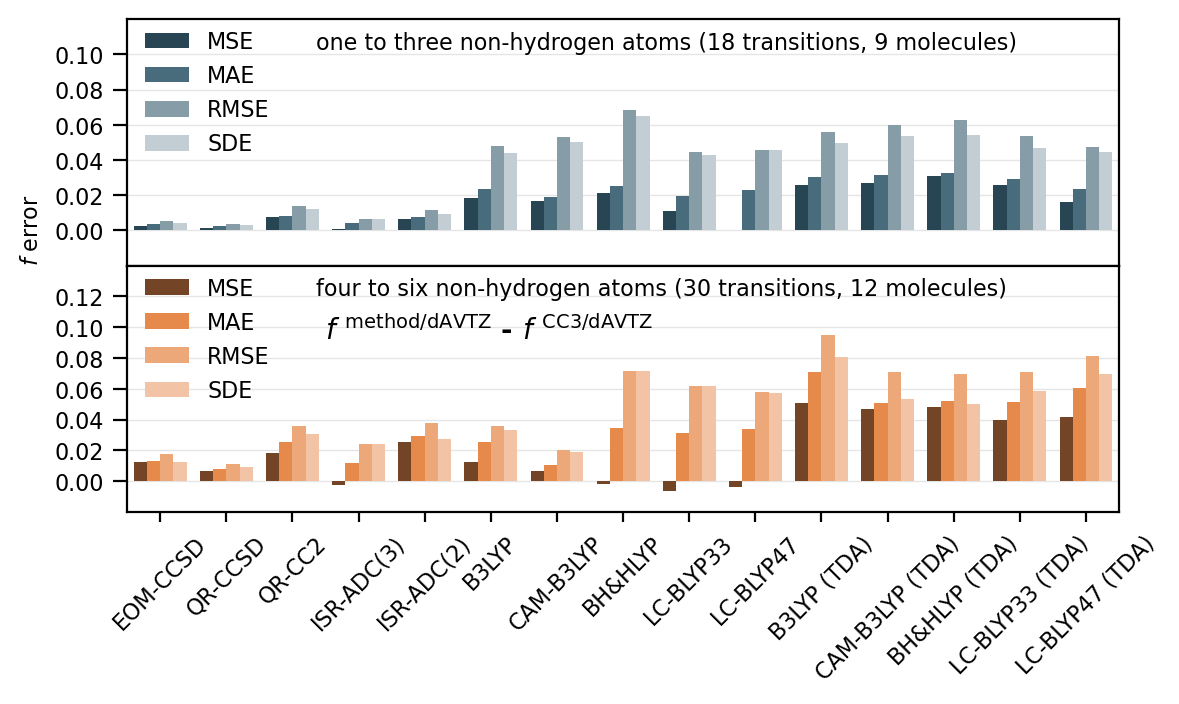}
   \caption{Comparison of errors in $f$ for all methods depending on the molecular size. Top: 1-3 non-hydrogen atom; bottom: 4-6 non-hydrogen atom. See the caption of Fig.~\ref{fig:Figure-5} for more details.}
            \label{fig:Figure-8}
    \end{figure*}

Upon investigating the effect of system size on the computed $f$ values (Fig.~\ref{fig:Figure-8}), three important trends emerge.
First, wavefunction methods perform in general worse for larger molecules, especially second-order methods, as upon increasing the system size the MAE of QR-CC2 [ISR-ADC(2)] grows from {\MAEsmallCCTW} ({\MAEsmallADCTW}) to {\MAEbigCCTW} ({\MAEbigADCTW}). 
Second, the impact of the TDA becomes more significant for larger systems: the TDA errors increase for larger molecules.
Third, with the exception of both B3LYP and CAM-B3LYP, the MAE, RMSE, and SDE of all functionals generally increase for larger molecules. Upon increasing the system size, the MAE of CAM-B3LYP decreases from {\MAEsmallTDDFTCAMBTHLYP} to {\MAEbigTDDFTCAMBTHLYP}. For B3LYP, the MAEs for smaller and larger molecules are almost identical. When we compare the change in SDE and RMSE, we see a stronger effect than for the MAE. In the case of CAM-B3LYP, for small molecules, one has SDE = {\STEsmallTDDFTCAMBTHLYP} and RMSE = {\RMSEsmallTDDFTCAMBTHLYP}, which both decrease in larger compounds with SDE = {\STEbigTDDFTCAMBTHLYP} and RMSE = {\RMSEbigTDDFTCAMBTHLYP}. For B3LYP, the change is less pronounced: for smaller systems one has SDE = {\STEsmallTDDFTBTHLYP} and RMSE = {\RMSEsmallTDDFTBTHLYP} and, for larger compounds, SDE = {\STEbigTDDFTBTHLYP} and RMSE = {\RMSEbigTDDFTBTHLYP}. The functional ranking, without TDA, thus changes dramatically for larger systems as B3LYP climbs up the performance ladder from the last place to the second one. Given that, as we mentioned before, in practical applications, the dyes used for ESA applications are rather large organic molecules, B3LYP does not seem to perform so badly after all. Most importantly, CAM-B3LYP, for larger systems, comes very close to wavefunction approaches and even outperforms ISR-ADC(2) (MAE = {\MAEbigADCTW}, RMSE = {\RMSEbigADCTW}, and SDE = {\STEbigADCTW}), CC2 (MAE = {\MAEbigCCTW}, RMSE = {\RMSEbigCCTW}, and SDE = {\STEbigCCTW}), and, in terms of MAE, EOM-CCSD (MAE = {\MAEbigEOMCCSD}, RMSE = {\RMSEbigEOMCCSD}, and SDE = {\STEbigEOMCCSD}).

\section{Conclusions}

We have defined highly accurate reference values of ${\Delta}E_{m, n}$ and $f$ values for a set of {\alltrans} excited-to-excited state ({\allstates} states) transitions in {\allmolecules} small- and medium-sized molecules using the QR-CC3/dAVTZ level of theory. We confirmed the validity of this benchmark level by studying a small set of transitions in {\smallsetmolecules} compact molecules by comparison with i) QR-CC3/dAVQZ, QR-CC3/AVQZ, QR-CC3/tAVTZ, and QR-CC3/tAVDZ values, ii) differences between $f$ obtained in different gauges, and iii) extrapolated FCI results. 

After verifying the quality of the QR-CC3/dAVTZ data, we explored the basis set effects for all transitions with three smaller basis sets: AVDZ, dAVDZ, and AVTZ. We found that, at the QR-CC3 level, augmentation with two sets of diffuse functions has a larger impact than $\zeta$-multiplicity. This is because the majority of states considered in our study present a Rydberg character. The basis set dependences at the QR-CCSD, QR-CC2, and ISR-ADC(2) levels of theory are consistent with QR-CC3. We further found that, for the majority of the transitions, the dAVDZ basis is sufficient when QR-CC3 is used. This smaller basis set can thus be employed to establish reference values when the QR-CC3/dAVTZ approach becomes computationally out of reach.

Next we compared $f$, ${\Delta}E_{{0,n}}$, and ${\Delta}E_{m, n}$ obtained with six different wavefunction approaches: QR-CC3 (reference), QR-CCSD, EOM-CCSD, QR-CC2, ISR-ADC(2), and ISR-ADC(3) as well as QR-TDDFT with five different XCFs (B3LYP, CAM-B3LYP, BH{\&}HLYP, LC-BLYP33, and LC-BLYP47) applying or not the TDA. We found that the MAE for $f$ values obtained using wavefunction approaches follows the trend: QR-CCSD < ISR-ADC(3) < EOM-CCSD < QR-CC2 < ISR-ADC(2). Among the functionals, CAM-B3LYP exhibited the best performance for ESA $f$ in terms of MAE, followed by LC-BLYP33, LC-BLYP47, BH\&HLYP, and B3LYP. The application of TDA is globally detrimental to the calculation of ESA oscillator strengths. Interestingly, ISR-ADC(3) and LC-BLYP47 (full TDDFT) give very accurate $f$ values for low-intensity transitions.




We further found that TD-DFT, without TDA, performs very well for valence-valence transitions and even outperforms some wavefunction approaches. However, since most bright ESA transitions studied here, include at least one Rydberg state, this does not impact the overall statistics. The implication of this outcome for practical applications on large dyes remains an open question.

After investigating the effect of increasing the molecular size on the computed $f$ values, we found that TDA introduces larger errors for larger molecules. Lastly, we discovered that the errors of CAM-B3LYP without TDA, become significantly smaller (and also more systematic) for the largest molecules of our set. For B3LYP, the errors become more systematic (but of similar magnitude) for larger systems. For the other methods, the trend was the opposite of CAM-B3LYP. The (CAM-)B3LYP trends are advantageous as real-life dyes are, in general, large organic molecules, as we mentioned before. Based on our analysis, we suggest relying on CAM-B3LYP for computing ESA oscillator strengths. 

We would like to stress that all calculations were performed in the gas phase. In fact, to the very best of our knowledge, a robust solvation model for QR methods has yet to be implemented in any widely available computational chemistry software. Therefore, the development and implementation of a suitable solvation model would be a valuable addition, particularly for applications to real-life systems.

\section*{Acknowledgments}
This work received financial support under the EUR LUMOMAT project and the Investment for the Future program ANR-18-EURE-0012, which supports the PhD thesis of JS. This research used resources from the GLICID Computing Facility (Ligerien Group for Intensive Distributed Computing 10.60487//glicid, Pays de la Loire, France). PFL and YD thank the European Research Council (ERC) under the European Union's Horizon 2020 research and innovation programme (Grant agreement No.~863481) for funding.
\section*{Supporting Information Available}
Geometries for all compounds. Additional results for all methods and basis sets.

\bibliography{Bibliography}

\end{document}